\documentclass[aps,amsmath,amssymb,preprint,showkeys,superscriptaddress,showpacs,floatfix,longbibliography]{revtex4-1}
\usepackage{graphicx}
\usepackage{dcolumn}
\usepackage[colorlinks=true,linkcolor=blue ,citecolor=blue,urlcolor=blue]{hyperref}
\usepackage{multirow}
\usepackage[usenames,dvipsnames]{xcolor}
\usepackage{soul}
\usepackage{braket}
\usepackage{bm}
\usepackage{nicefrac}
\usepackage{siunitx}
\usepackage{color}
\usepackage[utf8]{inputenc}
\usepackage{upgreek}
\usepackage{tikz}

\newcommand{\VEC}[1]{\mathbf{#1}}

\DeclareSIUnit\ML{ML}
\DeclareSIUnit\MLs{MLs}
\DeclareSIUnit\meVA{meV\angstrom^2}

\DeclareSIUnit\ML{ML}
\DeclareSIUnit\MLs{MLs}
\DeclareSIUnit\meVA{meV\angstrom^2}

\begin{document}

\title{Modelling spin waves in noncollinear antiferromagnets: spin-flop states, spin spirals, skyrmions and antiskyrmions}

\author{Flaviano José dos Santos}\email{f.dos.santos@fz-juelich.de}
\author{Manuel dos Santos Dias}
\author{Samir Lounis}
\affiliation{Peter Gr\"{u}nberg Institut and Institute for Advanced Simulation, Forschungszentrum J\"{u}lich and JARA, D-52425 J\"{u}lich, Germany}

\date{\today}

\begin{abstract}

Spin waves in antiferromagnetic materials have great potential for next-generation magnonic technologies.
However, their properties and their dependence on the type of ground-state antiferromagnetic structure are still open questions.
Here, we investigate theoretically spin waves in one- and two-dimensional model systems with a focus on noncollinear antiferromagnetic textures such as spin spirals and skyrmions of opposite topological charges.
We address in particular the nonreciprocal spin excitations recently measured in bulk antiferromagnet $\alpha$--$\textnormal{Cu}_2\textnormal{V}_2\textnormal{O}_7$ utilizing inelastic neutron scattering experiments [Phys.\ Rev.\ Lett.\ \textbf{119}, 047201 (2017)], where we help to characterize the nature of the detected spin-wave modes.
Furthermore, we discuss how the Dzyaloshinskii-Moriya interaction can lift the degeneracy of the spin-wave modes in antiferromagnets, resembling the electronic Rashba splitting.
We consider the spin-wave excitations in antiferromagnetic spin-spiral and skyrmion systems and discuss the features of their inelastic scattering spectra.
We demonstrate that antiskyrmions can be obtained with an isotropic Dzyaloshinskii-Moriya interaction in certain antiferromagnets.

\end{abstract}

\maketitle

\section{Introduction}

The search for new technologies that are faster and more energy-efficient than present-day electronics stimulated the development of spintronics~\cite{Zutic2004} and magnonics~\cite{chumak_magnon_2015}, which exploit magnetic degrees of freedom instead of only mobile charges as in conventional electronics.
The spin of the electron is central to spintronics, while magnonics builds upon spin waves or magnons, which are collective motions of magnetic moments.
Both traditionally involve ferromagnetic materials, but recently antiferromagnets were recognized to have potential advantages, which led to the development of antiferromagnetic spintronics~\cite{jungwirth_antiferromagnetic_2016,Wadley2016,Olejnik2018,Baltz2018} while antiferromagnetic magnonics is still in its infancy~\cite{Kampfrath2011,Daniels2018}.
Skyrmionics can be seen as an interesting crossover between these two fields~\cite{fert_skyrmions_2013,sampaio_nucleation_2013,crum_perpendicular_2015,Jiang2015,zhang_magnetic_2015,garcia-sanchez_skyrmion-based_2016,Fert2017,Back2020}.
Here, the key entity is the magnetic skyrmions~\cite{bogdanov_thermodynamically_1989,Bogdanov1994,Rossler2006,nagaosa_topological_2013} (topologically-quantized windings of the background magnetic structure), which can be very robust against perturbations and also highly mobile under relatively small applied currents.
Skyrmions have been theoretically studied in antiferromagnetic materials~\cite{barker_static_2016,Velkov2016,Keesman2016,Zhang2016a,Goebel2017b,Kravchuk2019,diaz_topological_2019}, but have not yet been experimentally discovered.

Broadening the applications of antiferromagnetic materials, for instance, to advance antiferromagnetic magnonics, requires understanding the properties of their spin waves.
The basic quantities are the spin-wave energy, how it relates to the wavevector, and what kind of precession of the magnetic moments takes place.
The pioneering works of Kittel and Keffer~\cite{Kittel1951,Keffer1952} explain antiferromagnetic resonance (zero wavevector) in collinear antiferromagnets, and how it depends on the internal magnetic anisotropy and the external magnetic field.
For larger wavevectors, the spin-wave energies are controlled by the magnetic exchange interaction and the Dzyaloshinskii-Moriya interaction (DMI)~\cite{dzyaloshinsky_thermodynamic_1958,moriya_anisotropic_1960}.
The DMI is a chiral coupling that arises from the relativistic spin-orbit interaction and was originally proposed to explain weak ferromagnetism in collinear antiferromagnetic materials.
It can also lead to noncollinear antiferromagnetic structures, such as spin spirals and potentially antiferromagnetic skyrmions~\cite{Bogdanov2002,Fishman2019}.
In ferromagnetic materials, the DMI leads to the nonreciprocity of the spin-wave dispersion \cite{udvardi_chiral_2009,costa_spin-orbit_2010,zakeri_asymmetric_2010,dos_santos_nonreciprocity_2020}, so that the spin-wave energy is different for wavevectors of equal length and opposite direction.
A similar effect was recently observed in an antiferromagnetic material~\cite{gitgeatpong_nonreciprocal_2017}.
While spin waves are now well-understood in skyrmion-hosting ferromagnetic materials~\cite{garst_collective_2017}, only some theoretical studies report on their antiferromagnetic counterparts~\cite{Kravchuk2019,diaz_topological_2019}, as they remain to be discovered.

A complete experimental characterization of the spin-wave modes in an antiferromagnetic material is challenging, especially detecting what kind of precession is associated with each mode.
Antiferromagnetic resonance can be accessed by broadband spectroscopy~\cite{Kampfrath2011,Nemec2018}, while inelastic neutron scattering can survey the spin-wave spectrum across the whole Brillouin zone, as shown for the antiferromagnetic spin-spiral system Ba$_2$CuGe$_2$O$_7$~\cite{Zheludev1999}.
Polarized neutron scattering experiments give access to the information about the precession but are very difficult to perform.
Resonant inelastic x-ray scattering is a new technique that holds promise for detecting spin waves in bulk materials~\cite{Kim2012a} and is applicable down to a single layer~\cite{Dean2012}.
Recently, we presented a theoretical analysis of spin-resolved electron-energy-loss spectroscopy (SREELS)~\cite{dos_santos_spin-resolved_2018}, which is a proposed extension of an existing surface-sensitive experimental method~\cite{plihal_spin_1999,vollmer_spin-polarized_2003,michel_spin_2015,dos_santos_first-principles_2017}, and explained how the polarization analysis can be used to understand the nature of the spin-wave precession.
Employing SREELS, one has access to various spin-scattering channels, where the scattered electrons experience processes of spin-flip nature or not.
In contrast to collinear magnets, where only spin-flip processes are responsible for the emission of spin waves, non-spin-flip processes can generate spin-excitations in noncollinear materials.
These general concepts are also applicable to other spectroscopies.

In this article, we study the inelastic scattering spectra of various noncollinear antiferromagnetic spin structures in one and two dimensions using atomistic model systems.
In particular, we address the nonreciprocity induced by the combined action of the DMI and an external magnetic field.
We compare our results to inelastic neutron scattering measurements on a bulk antiferromagnet, $\alpha-\textnormal{Cu}_2\textnormal{V}_2\textnormal{O}_7$~\cite{gitgeatpong_nonreciprocal_2017}, providing a novel understanding of the experimental data.
In two dimensions, we explore antiferromagnetic systems with the $C_{4v}$ and $C_{2v}$ symmetries.
We showed that DMI in these systems can induce a spin-wave Rashba-like effect characterized by a linear-angular-momenta locking.
Furthermore, we demonstrate that antiskyrmions are natural skyrmionic occurrences in antiferromagnetic p($2\times1$) materials, even when the system has the $C_{4v}$ symmetries.
Finally, we calculate the inelastic scattering spectra of antiferromagnetic skyrmion and antiskyrmion lattices, identifying the spectral signature of its characteristic modes.

\section{Theoretical framework and model systems}
\label{sec:theoretical_framework}
    
We employ the generalized Heisenberg model for spins $\VEC S_i$ (we take $S = 1$) on a lattice, and we measure lengths in units of the nearest-neighbor distance $a = 1$. 
The corresponding Hamiltonian restricted to nearest-neighbor interactions reads
\begin{equation}\label{eq:Hamiltonian}
  \mathcal{H} = \sum_{\langle i,j \rangle} J_{ij}\,\VEC S_i \cdot \VEC S_j
  - \sum_{\langle i,j \rangle} \VEC{D}_{ij} \cdot \big(\VEC S_i \times \VEC S_j\big)
  - K \sum_i \big(S^z_i\big)^2 - \sum_i \VEC{B} \cdot \VEC{S}_i \quad .
\end{equation}
Here the exchange interaction $J_{ij}$ is taken to be uniform and antiferromagnetic $J > 0$, except in one of the considered models.
The Dzyaloshinskii-Moriya interaction $\VEC{D}_{ij} = D\,\hat{\VEC n}_{ij}$ is taken to be uniform in magnitude $D$, with the direction for each bond given by $\hat{\VEC n}_{ij}$.
We also include a uniaxial magnetic anisotropy with $K > 0$ and easy-axis along $z$, and the external uniform magnetic field $\VEC{B}$ with magnitude $B$, also along $z$ for most of the models.
The brackets indicate a sum over the nearest-neighbor pairs.

For the various investigated cases, we extract the most stable magnetic configuration either by analytical means or by using atomistic-spin-dynamics simulations by solving the Landau–Lifshitz–Gilbert equation with the \textit{Spirit} code~\cite{muller_spirit:_2019}.
For that, we used a supercell approach with periodic boundary conditions.
Once the ground state or a metastable state is found, we compute the adiabatic spin-wave modes and the corresponding inelastic scattering spectrum, based on time-dependent perturbation theory.
The associated theoretical framework was presented in Refs.~\cite{dos_santos_first-principles_2017,dos_santos_spin-resolved_2018}.
Although we have access to several distinct scattering channels, in this work we always report results for the total inelastic scattering spectrum due to spin waves (the sum over all scattering channels), as one would measure in an experiment with an unpolarized scattering experiment.

An important property of the spin-wave quantum is its angular momentum.
In ferromagnets, spin waves have quantized angular momenta oriented antiparallel to the spins forming the background magnetization.
Meanwhile, noncollinear systems can host spin-wave modes with nonquantized angular momenta pointing in various directions, or even of vanishing magnitude~\cite{dos_santos_spin-resolved_2018}.
Spin waves with finite angular momenta correspond to excitations with circular polarization and are also called rotational or gyroscopic modes.
In contrast, modes with vanishing angular momenta are linearly polarized and can be seen as longitudinal excitations.
Recently, we have shown that the angular momentum of a spin wave is connected to its handedness (chirality), determining how the spin-wave properties respond to the DMI~\cite{dos_santos_nonreciprocity_2020}.
Experimentally, this angular momentum can be measured via spin-resolved inelastic scattering spectroscopy.
That is because the angular momentum determines in which scattering channels a given mode may be observed~\cite{dos_santos_spin-resolved_2018}.
We used the theoretical SREELS spectra to determine the spin-wave angular momenta in this work.

\section{Antiferromagnetism in a spin chain}\label{sec:1d_chain}
To set the stage, we first study the case of one-dimensional antiferromagnetic spin chains.

\subsection{Collinear antiferromagnetic chains: The effect of the DMI and the magnetic field on the spin waves}
\label{sec:effect_of_DMI_rashba}

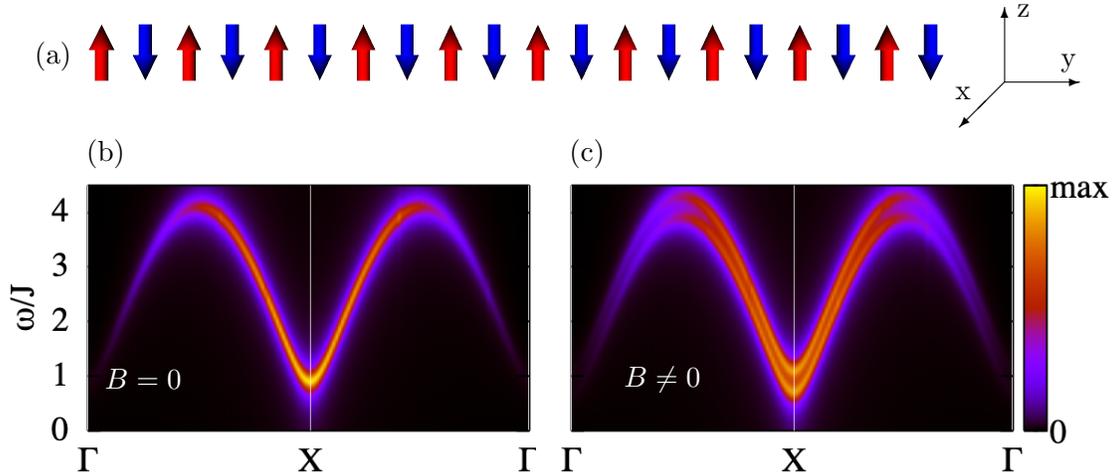
\begin{figure}[t]
\setlength{\unitlength}{1cm} 
\newcommand{\boxsize}{0.3}

  \begin{picture}(15,6.5)

    \put(0,-1.2){
        \put(1, 6){ \includegraphics[width=12 cm,trim={0 30cm 0 30cm},clip=true]{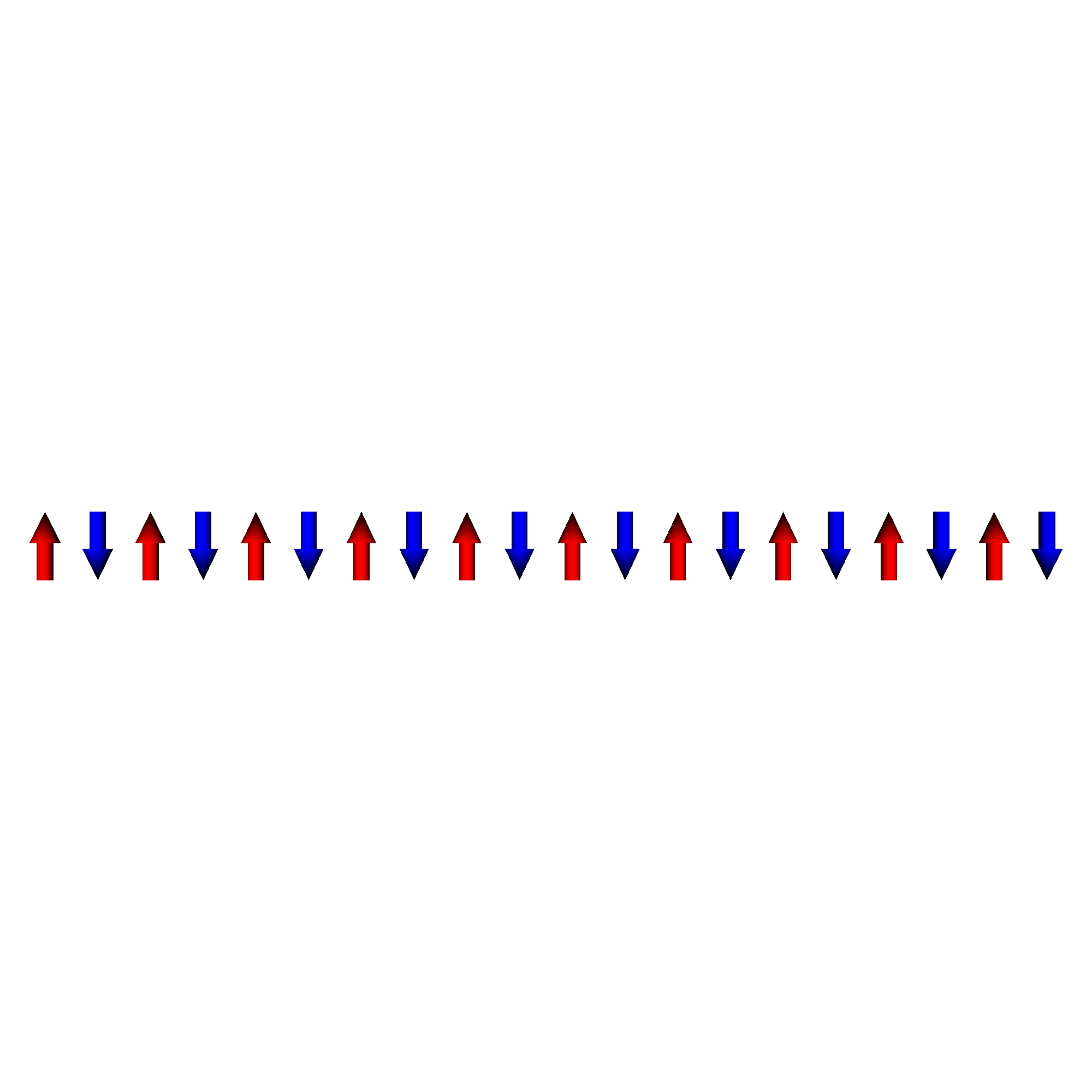} }
        
        \put(0.7,6.7){ \makebox(\boxsize,\boxsize){(a)} } 
        
        \newcommand{\axisposX}{13.5}
        \newcommand{\axisposY}{6.5}
        \put(\axisposX,\axisposY){ \vector(1,0){1}}
        \put(\axisposX,\axisposY){ \vector(0,1){1}}
        \put(\axisposX,\axisposY){ \vector(-1,-1){0.6}}
        \put(14.2,6.6){ \makebox(\boxsize,\boxsize){y} } 
        \put(13.6,7.3){ \makebox(\boxsize,\boxsize){z} } 
        \put(12.8,6.2){ \makebox(\boxsize,\boxsize){x} } 
    }
    
    \put(0, 0){ \includegraphics[height=4.5 cm,trim={0 1cm 3.55cm 0},clip=true]{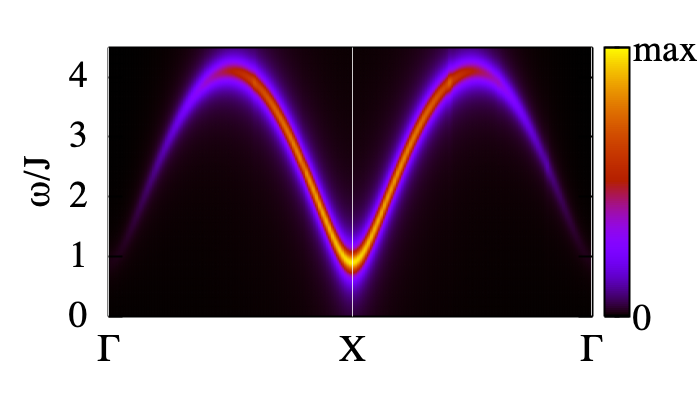} }
    
    \put(7.5, 0){ \includegraphics[height=4.5 cm,trim={3.1cm 1cm 0 0},clip=true]{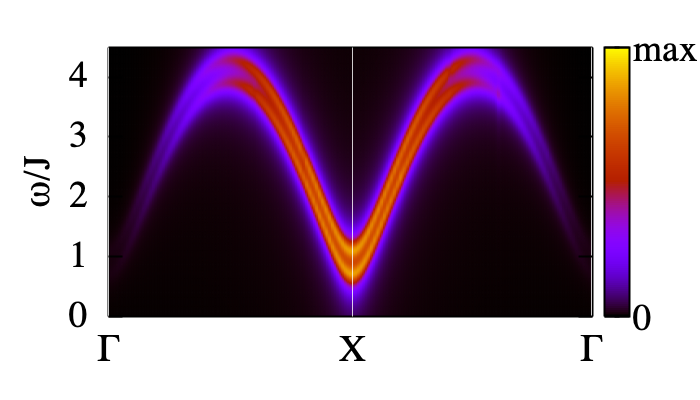} }

    \put(1.4,4.2){ \makebox(\boxsize,\boxsize){(b)} } 
    \put(7.8,4.2){ \makebox(\boxsize,\boxsize){(c)} } 

    \color{white}
    \put(1.9,1.2){ \makebox(\boxsize,\boxsize){$B=0$} } 
    \put(8.8,1.2){ \makebox(\boxsize,\boxsize){$B\neq0$} }
    
  \end{picture}

  \caption{\label{fig:Fig_AF}
  Inelastic scattering spectrum of a collinear antiferromagnetic spin chain.
  (a) The ground state, where the spins align antiparallel among neighbors. 
  (b) Inelastic scattering spectrum for the antiferromagnetic spin chain depicted in (a).
  The single spin-wave mode dispersing away from $X$ is doubly-degenerate and the excitation gap is due to the magnetic anisotropy.
  (c) A magnetic field is applied along the $z$--axis, which breaks the degeneracy of the two modes. 
  Model parameters: $D=0$, $K=0.05 J$, $B=0.2J$.
  }
\end{figure}

When $B$ and $D$ are zero, i.e., considering only the magnetic exchange interaction and the magnetocrystalline anisotropy, the ground state of the system corresponds to a collinear alignment of the spins along the anisotropy easy-axis $z$.
The nearest-neighbor spins are antiparallel to each other, as shown in Fig.~\ref{fig:Fig_AF}(a), the inelastic scattering spectrum shows a single mode even though one could expect two, given that there are two magnetic sublattices (one for up and the other for down spins), see Fig.~\ref{fig:Fig_AF}(b).
This spin-wave mode is doubly-degenerate, having its lowest excitation energy at $\Gamma$ (with vanishing scattering intensity) and at the Brillouin zone border, $\mathrm{X} = \pi$, with an excitation gap opened by the magnetic anisotropy.
The degeneracy of the two modes is lifted by an external magnetic field parallel to the easy-axis.
As the modes have opposite angular momenta (they can be measured individually with spin-resolved inelastic scattering spectroscopy~\cite{dos_santos_spin-resolved_2018}),
the magnetic field raises the energy of one mode while lowers the energy of the other, as seen in Fig.~\ref{fig:Fig_AF}(c).

\begin{figure}[t]
\setlength{\unitlength}{1cm} 
\newcommand{\boxsize}{0.3}

  \begin{picture}(15,4.5)
    
    \put(0, 0){ \includegraphics[width=14.8 cm,trim={3cm 1cm 1cm 0},clip=true]{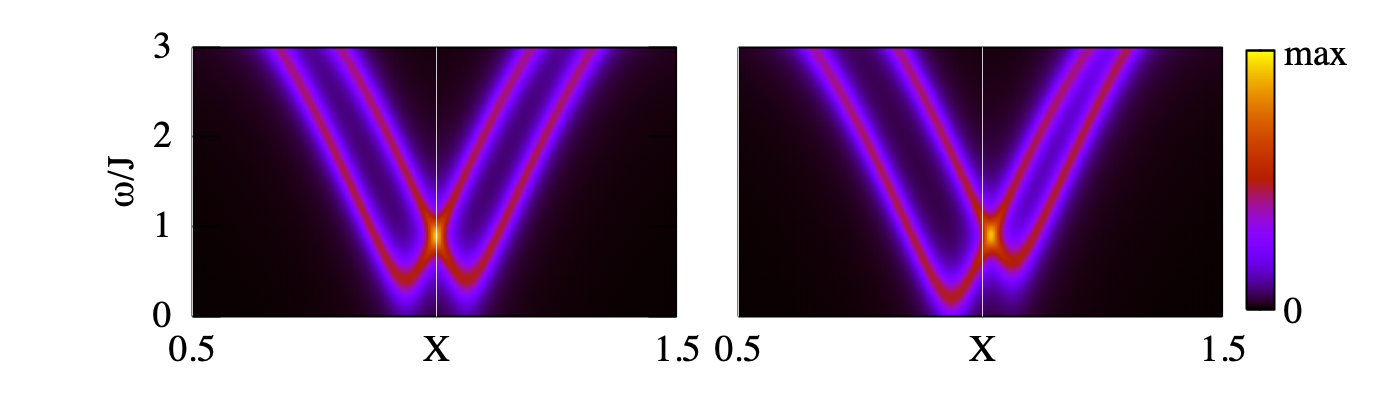} }

    \put(1.25,4.1){ \makebox(\boxsize,\boxsize){(a)} } 
    \put(7.55,4.1){ \makebox(\boxsize,\boxsize){(b)} }
    
    \color{white}
    \put(2.0,1.3){ \makebox(\boxsize,\boxsize){$B=0$} } 
    \put(8.1,1.3){ \makebox(\boxsize,\boxsize){$B\neq0$} } 

  \end{picture}

  \caption{\label{fig:Fig_AF_DMI_B}
  Inelastic scattering spectrum of a collinear antiferromagnetic spin chain with DMI.
  (a) The Dzyaloshinskii-Moriya vectors pointing along the easy-axis break the spin-wave mode degeneracy.
  The dispersion curves shift in opposite directions.
  (b) When a magnetic field is applied ($B=0.2J$),
  the spectrum becomes nonreciprocal.
  Despite the DMI, the system has the same collinear ground state shown in Fig.~\ref{fig:Fig_AF}(a), stabilized by the magnetic anisotropy.
  Model parameters: $D=0.2J$, $K=0.05J$, $B=0.2J$.
  The wavevector is given in units of $\pi/a$ (we set the lattice spacing $a = 1$).
  }
\end{figure}
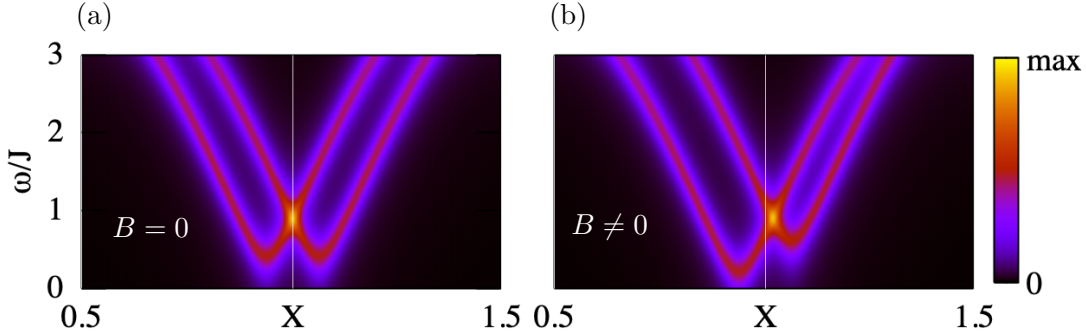

In the absence of an external magnetic field, the Dzyaloshinskii-Moriya interaction can also lift the degeneracy.
Introducing the DMI vectors collinear with the easy-axis but with magnitude below the critical value 
$D_\textnormal{c} = \sqrt{\left(J + \frac{K}{4}\right) K}$, see Appendix~\ref{Apx:Stiff_spiral_tensors}, preserves the collinear antiferromagnetic ground state.
The DMI splits the modes apart shifting their dispersion curves in opposite directions in the reciprocal space, as shown in Fig.~\ref{fig:Fig_AF_DMI_B}(a).
These spin-wave modes have their minimum excitation energies at $\mathrm{X} \pm q$, which were symmetrically shifted away from the Brillouin zone boundary.
This happens because the spiralization induced by the spin waves is energetically favored by the DMI, such that $q = \arctan(D/J)$.
Also, the energy gap is now smaller than in the absence of DMI, and it closes completely for the critical DMI magnitude, $D_\textnormal{c}$.
This phenomenon is analogous to the Rashba effect~\cite{bychkov_properties_1984}, where electrons acquire a finite group velocity at zero wavevector due to the spin-orbit coupling.
Furthermore, in the electronic Rashba effect, electrons propagating to opposite directions with the same wavelength have opposite spins.
Similarly, the two spin-wave modes in Fig.~\ref{fig:Fig_AF_DMI_B} have opposite angular momenta along the $z$--axis, thus perpendicular to the propagation direction $y$.

Finally, we can turn on the external magnetic field parallel to the easy-axis in the presence of the DMI.
Again, one of the curves is raised in energy, while the other is lowered.
This causes a break of the symmetry in the reciprocal space, which is a phenomenon known as nonreciprocity, as observed in Fig.~\ref{fig:Fig_AF_DMI_B}(b).
Note that the isolated action of either the magnetic field or the DMI preserves the reciprocity of the spectrum.

\subsection{Antiferromagnetic spin spirals}\label{sec:ass}

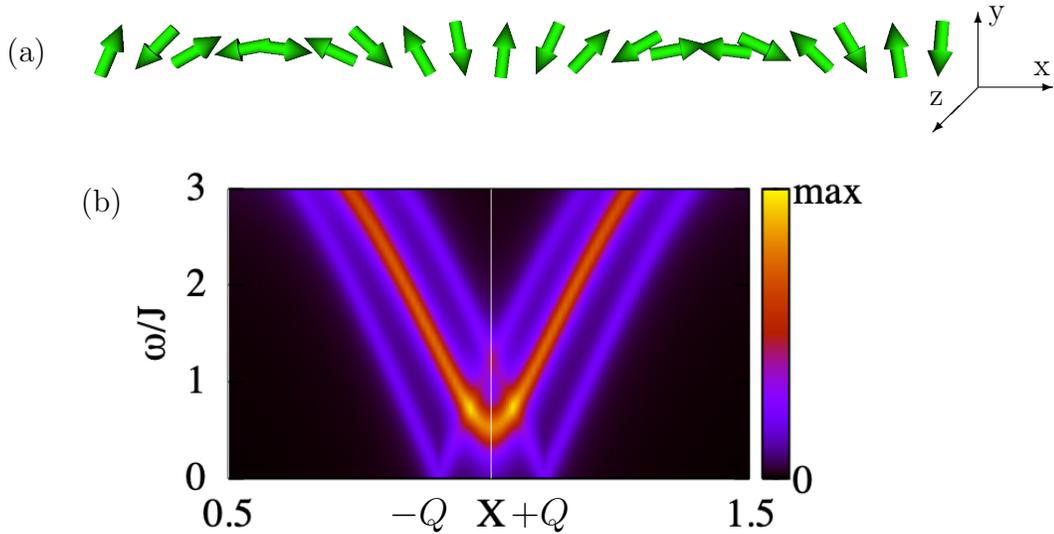
\begin{figure}[!ht]
\setlength{\unitlength}{1cm} 
\newcommand{\boxsize}{0.3}

  \begin{picture}(14,8)
    \put(1.5, 6){ \includegraphics[width=12 cm,trim={0 30cm 0 30cm},clip=true]{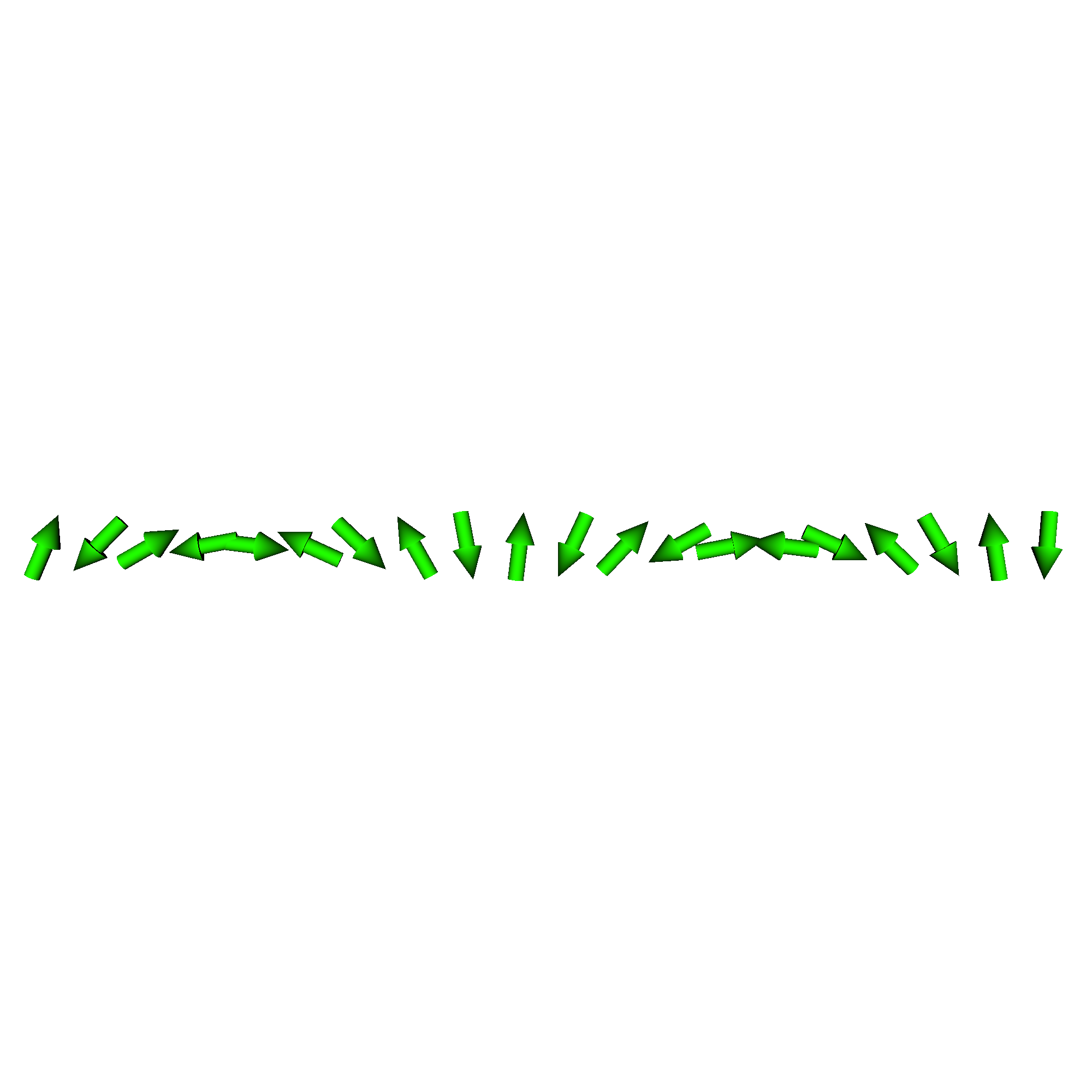} }
    \put(2, 0){ \includegraphics[width=10 cm,trim={0 0 0 0},clip=true]{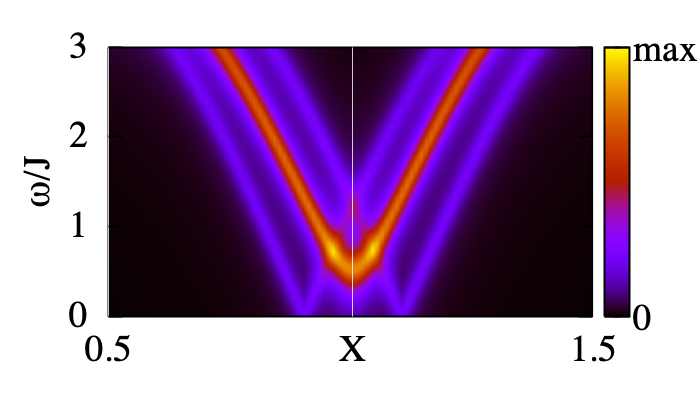} }

    \large
    \put(5.89,0.55){ \makebox(\boxsize,\boxsize){$-Q$} } 
    \put(7.50,0.55){ \makebox(\boxsize,\boxsize){$+Q$} } 
    \normalsize
    
    \put(0.7,6.7){ \makebox(\boxsize,\boxsize){(a)} } 
    \put(1.7,4.7){ \makebox(\boxsize,\boxsize){(b)} } 

    \newcommand{\axisposX}{13.5}
    \newcommand{\axisposY}{6.4}
    \put(\axisposX,\axisposY){ \vector(1,0){1}}
    \put(\axisposX,\axisposY){ \vector(0,1){1}}
    \put(\axisposX,\axisposY){ \vector(-1,-1){0.6}}
    \put(14.2,6.5){ \makebox(\boxsize,\boxsize){x} } 
    \put(13.6,7.2){ \makebox(\boxsize,\boxsize){y} } 
    \put(12.8,6.1){ \makebox(\boxsize,\boxsize){z} } 

  \end{picture}

  \caption{\label{fig:Fig_AF_SP}
  Inelastic scattering spectrum of an antiferromagnetic spiral spin chain.
  (a) The ground state generated by a spin-flop transition induced by an external magnetic field along $z$.
  The spins lie in the plane perpendicular to the applied field forming an antiferromagnetic spin spiral.
  A small component of each spin still points along $z$ what makes it a conical spin spiral.
  (b) Spin-wave scattering spectrum of the antiferromagnetic spin spiral depicted in (a).
  The spectrum is symmetric and formed by an intense gapped mode centered at $\mathrm{X}$, enclosed by two faint gapless modes dispersing away from $\pm Q = \mathrm{X} \pm q$, where $q$ is the wavevector of the spiral.
  Model parameters: $D=0.2J$, $K=0.05J$, $B=0.8J$.
  }
\end{figure}

In the current situation, we have an external magnetic field which is parallel to half of the spins and antiparallel to the other half.
Therefore, increasing the magnetic field does not affect the total energy of the collinear AFM state shown in Fig.~\ref{fig:Fig_AF}(a).
However, when the magnetic field is large enough, the system undergoes a spin-flop transition, where the spins are mostly perpendicular to the easy-axis but with a small component parallel to the field.
Furthermore, an antiferromagnetic spin spiral can form with a rotational axis parallel to the easy-axis such as to gain energy from the DMI, see Fig.~\ref{fig:Fig_AF_SP}(a).
This new state is unfavored by the magnetic anisotropy, but the loss is compensated by the Zeeman and the DMI energy gains.
The pitch of the antiferromagnetic spin spiral is given by $q=\arctan(D/J)$ just as for ferromagnetic systems, as shown in Appendix~\ref{Apx:Stiff_spiral_tensors}.
The inelastic scattering spectrum of this new state becomes reciprocal again.
Despite the stronger magnetic field, which previously was causing the nonreciprocity in combination with the DMI, the inelastic scattering spectrum of this new state is reciprocal.
The signal is formed by a central feature surrounded by two side modes of lower intensity, whose energy minima occurs at $\pm Q = X \pm q$.

\begin{figure}[!ht]
\setlength{\unitlength}{1cm} 
\newcommand{\boxsize}{0.3}

  \begin{picture}(15,6)
    \put(0, 0){ \includegraphics[width=14.5 cm,trim={0 0 0 0},clip=true]{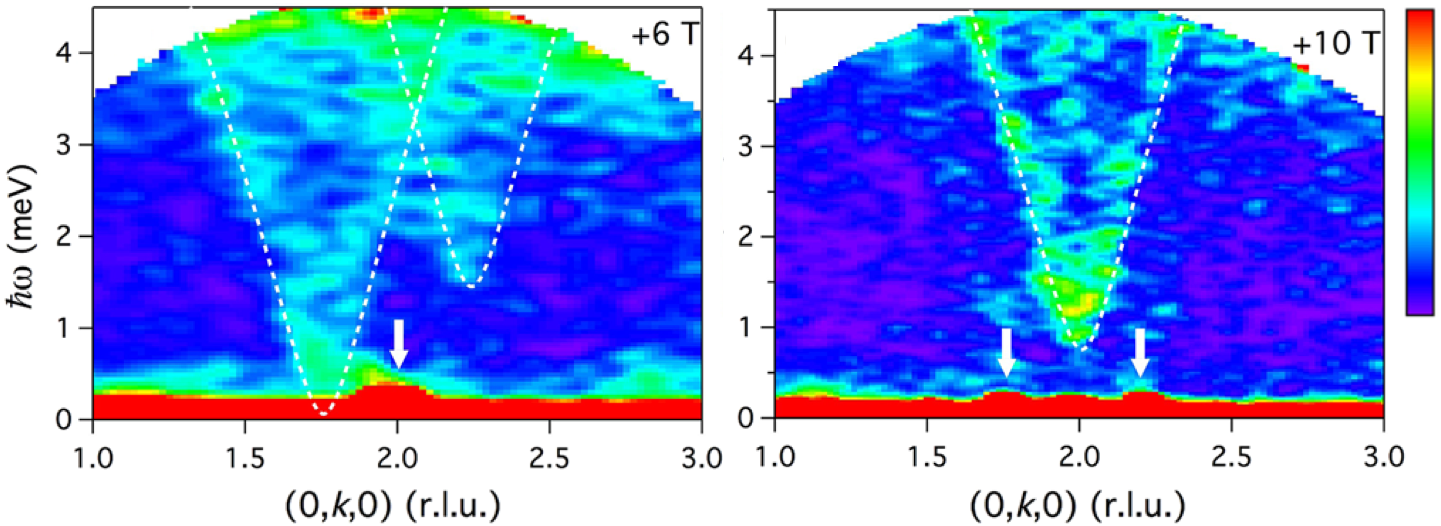} }

    \put(1.1,4.7){ \makebox(\boxsize,\boxsize){(a)} } 
    \put(8.0,4.7){ \makebox(\boxsize,\boxsize){(b)} } 
  \end{picture}

  \caption{\label{fig:Fig_exp}
  Inelastic neutron scattering spectra of bulk $\alpha$--$\textnormal{Cu}_2\textnormal{V}_2\textnormal{O}_7$.
  (a) Nonreciprocal scattering spectrum due to an external magnetic field, $B = \SI{6}{\tesla}$.
  The system is collinear antiferromagnetic.
  (b) For a high field, $B = \SI{10}{\tesla}$, the system undergoes a spin-flop transition into a spin spiral and the spectrum becomes symmetric.
  The arrows denote the magnetic Bragg peaks.
  Figure reprinted with permission from Gitgeatpong, G.\ \textit{et al.}, Phys.\ Rev.\ Lett.\ \textbf{119}, 047201 (2017) \href{https://doi.org/10.1103/PhysRevLett.119.047201}{(https://doi.org/10.1103/PhysRevLett.119.047201)}, Ref.~\cite{gitgeatpong_nonreciprocal_2017}. Copyright (2020) by the American Physical Society.
  }
\end{figure}

In Ref.~\cite{gitgeatpong_nonreciprocal_2017}, Gitgeatpong \textit{et al.} measured for the first time the nonreciprocity of spin waves in antiferromagnets by performing inelastic neutron scattering experiments on the bulk antiferromagnet $\alpha$--$\textnormal{Cu}_2\textnormal{V}_2\textnormal{O}_7$.
The spin-wave physics of this compound is analogous to the model just described.
The nonreciprocity is observed by applying an external magnetic field as shown in Fig.~\ref{fig:Fig_exp}(a), which resembles Fig.~\ref{fig:Fig_AF_SP}(a).
Our results establish a perfect parallel with their measurements.

We now present our understanding of these experimental results.
A spin spiral hosts three universal spin-wave modes~\cite{dos_santos_spin-resolved_2018}, instead of two for collinear two-sublattice antiferromagnets.
Two of the spin-wave modes of the spiral have dispersion curves with minima in the wavevector of the spiral $\pm Q$ and are rotational modes with equal and opposite net angular momenta.
The third mode is symmetric around a high-symmetry point ($\Gamma$ or $\mathrm{X}$), and it has no net angular momentum but corresponds to a longitudinal oscillation of the net magnetization, which is generated perpendicularly to the plane of rotation of the spiral (see a related discussion in Ref.~\cite{dos_santos_spin-resolved_2018}).
This longitudinal (linear polarized) mode is the one responsible for the high-intensity feature in the inelastic scattering spectrum, as obtained theoretically in Fig.~\ref{fig:Fig_AF_SP}(b), and measured in Fig.~\ref{fig:Fig_exp}(b).
Furthermore, our theoretical calculation enlightens the existence of two weaker features in the inelastic scattering spectrum, see Fig.~\ref{fig:Fig_AF_SP}(b).
These two modes have energy minima at the magnetic Bragg peaks of the antiferromagnetic spin spiral, which are given by the spiral wavevector $\pm Q$.
Therefore, the shifts of the minima out of the high-symmetry point are only related to the DMI indirectly through $Q$.
Although signatures of these two feeble branches also appear in the experimental data shown in Fig.~\ref{fig:Fig_AF_SP}(b), the insufficient counts and lack of theoretical support probably led the authors of Ref.~\cite{gitgeatpong_nonreciprocal_2017} to leave them unremarked.

To close this section, we remark that the nonreciprocity of spin waves in noncollinear systems is discussed at length in Ref.~\cite{santos2020nonreciprocity}, where we present a general theory of how to detect asymmetries in the inelastic scattering spectrum due to the DMI.

\section{Two-dimensional antiferromagnets}

In the previous section, we have considered a one-dimensional antiferromagnetic spin chain, which allowed us to study the spin-wave Rashba effect due to the Dzyaloshinskii-Moriya interaction as well as the reciprocal-symmetry breaking in response to an applied magnetic field.
Now, we place our focus on the properties of spin waves in two-dimensional antiferromagnetic systems, which have the minimal dimension to host antiferromagnetic skyrmionic structures as discussed, for example, in Refs.~\cite{barker_static_2016,diaz_topological_2019}.
First, we consider the formation of antiferromagnetic spin spirals, and subsequently the occurrence of antiferromagnetic skyrmions and antiskyrmions.

\subsection{Antiferromagnetic spin spirals and Rashba spin locking}

\begin{figure}[!ht]
\setlength{\unitlength}{1cm} 

\newcommand{\figheight}{3.5}

  \begin{picture}(14, 10 )
    
    \put(0.1,  \figheight){ \includegraphics[height=\figheight cm, trim={5cm 11cm 5cm 30cm},clip=true]{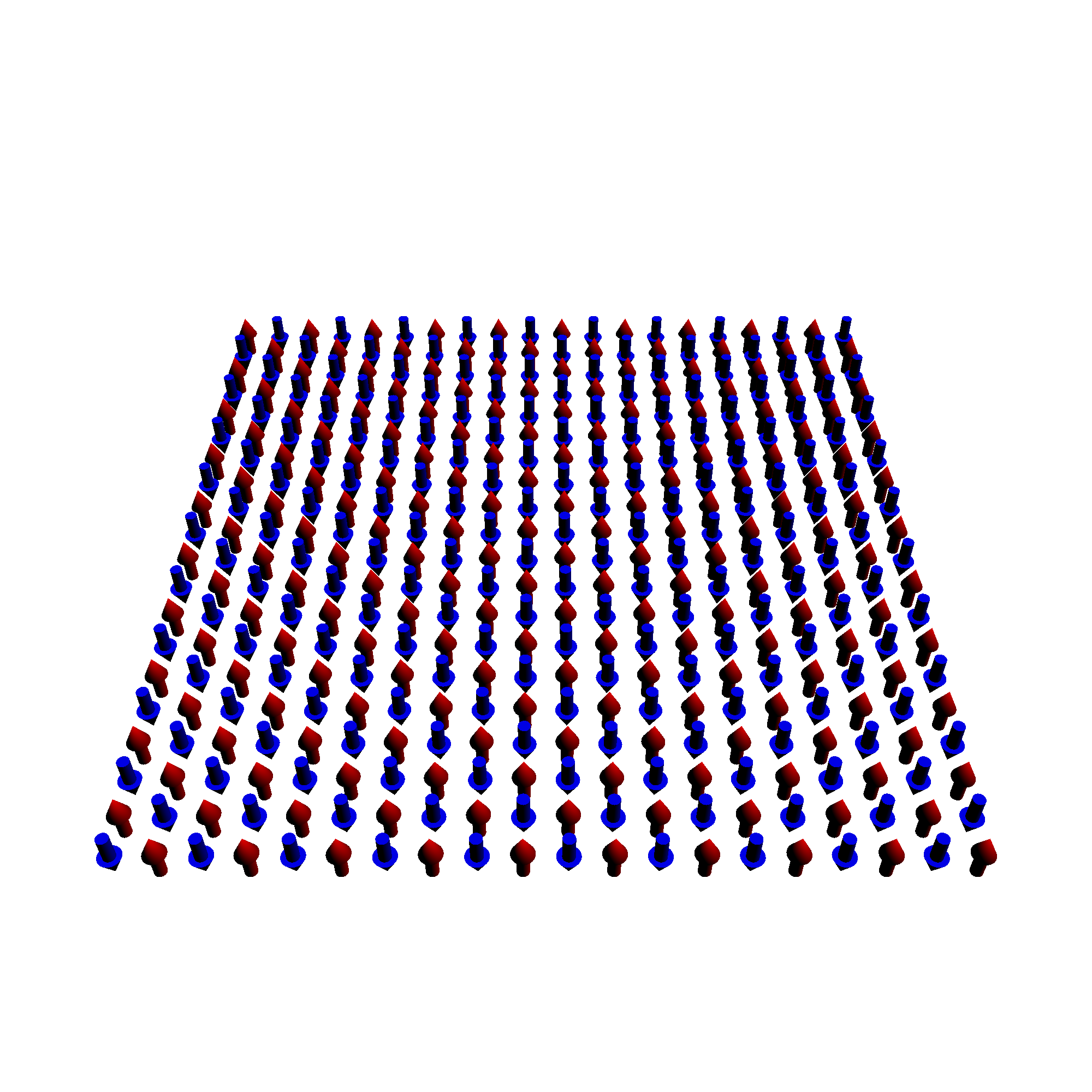} }
    \put(7.5,\figheight){ \includegraphics[height=\figheight cm, trim={5cm 11cm 5cm 30cm},clip=true]{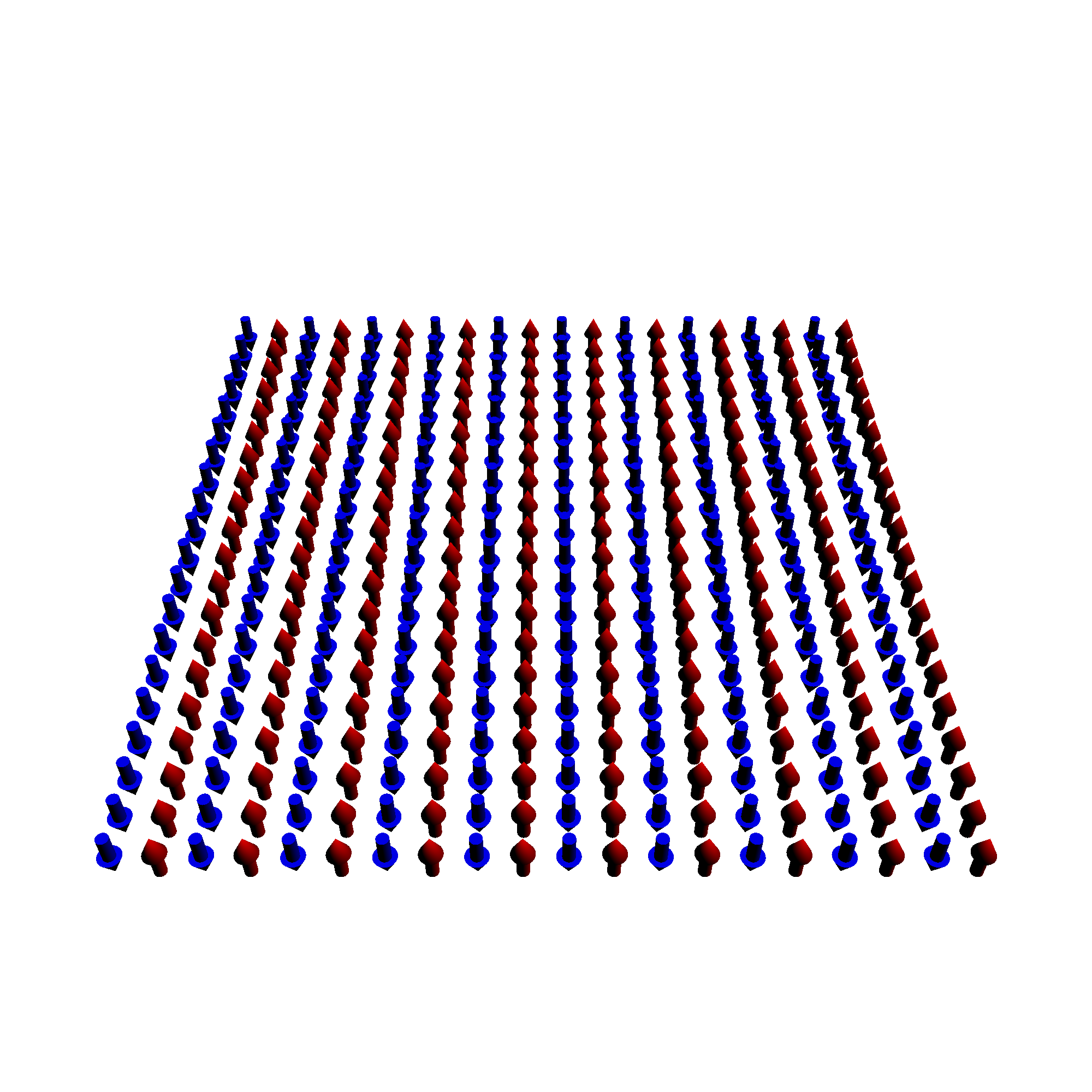} }

    \put(0.1,  0){ \includegraphics[height=\figheight cm, trim={5cm 11cm 5cm 30cm},clip=true]{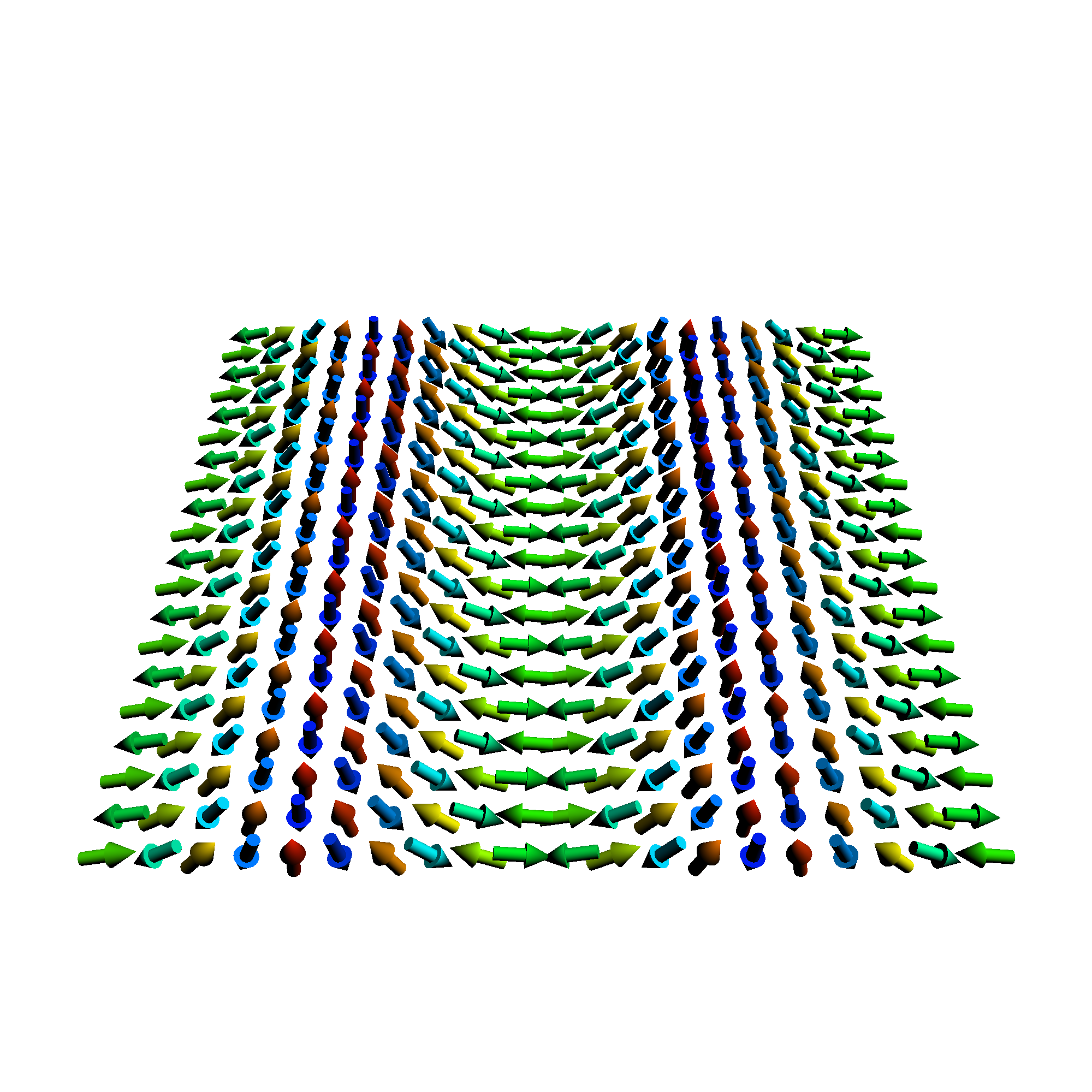} }
    \put(7.5,0){ \includegraphics[height=\figheight cm, trim={5cm 11cm 5cm 30cm},clip=true]{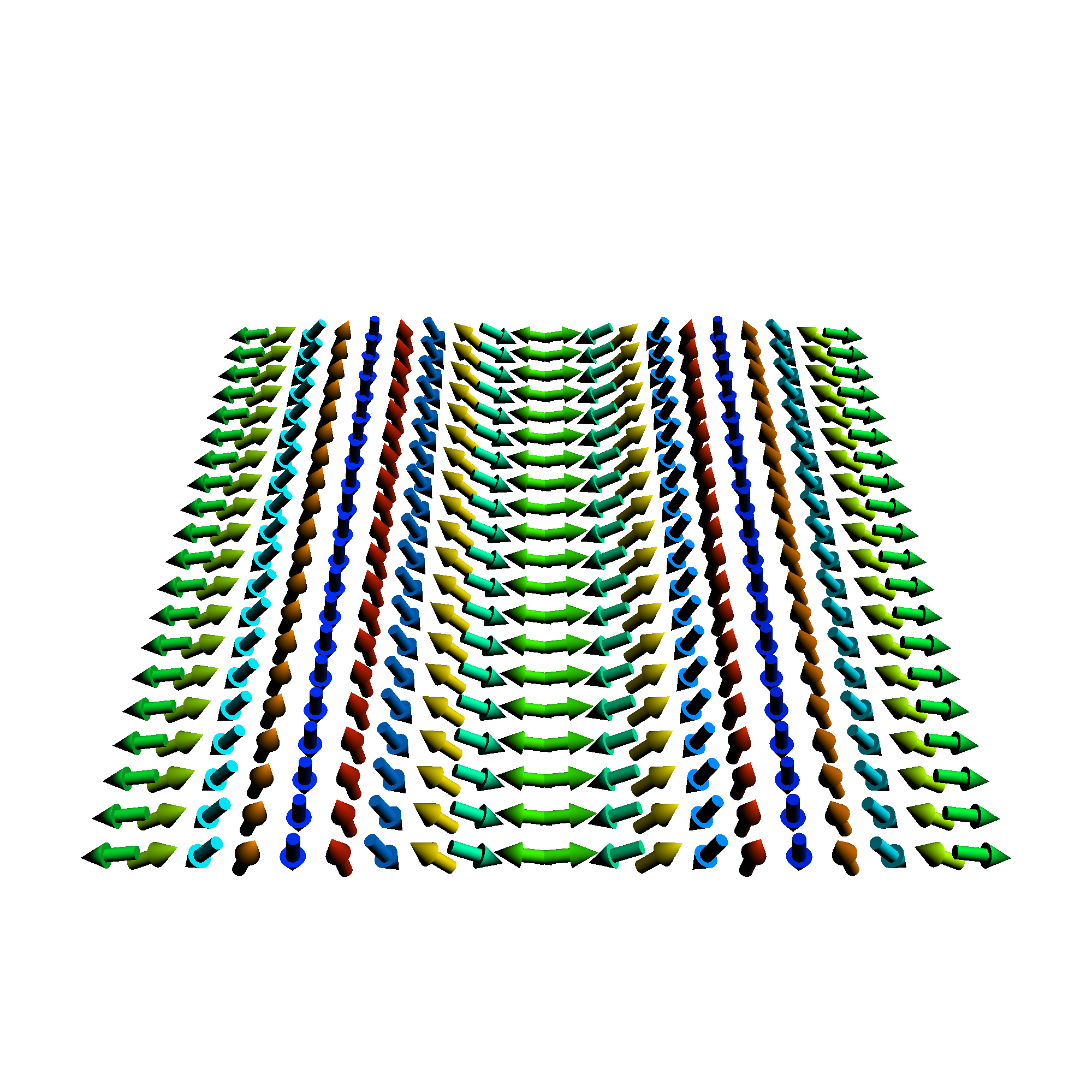} }
    
    \put(0, 7){
      \put(1.4,2.2){ \makebox(1,1){ (a) } } 
      \put(8.7,2.2){ \makebox(1,1){ (b) } }
      
      \put(1.7, 0 ){ \includegraphics[height=4 cm, trim={0cm 0cm 0cm 0cm},clip=true]{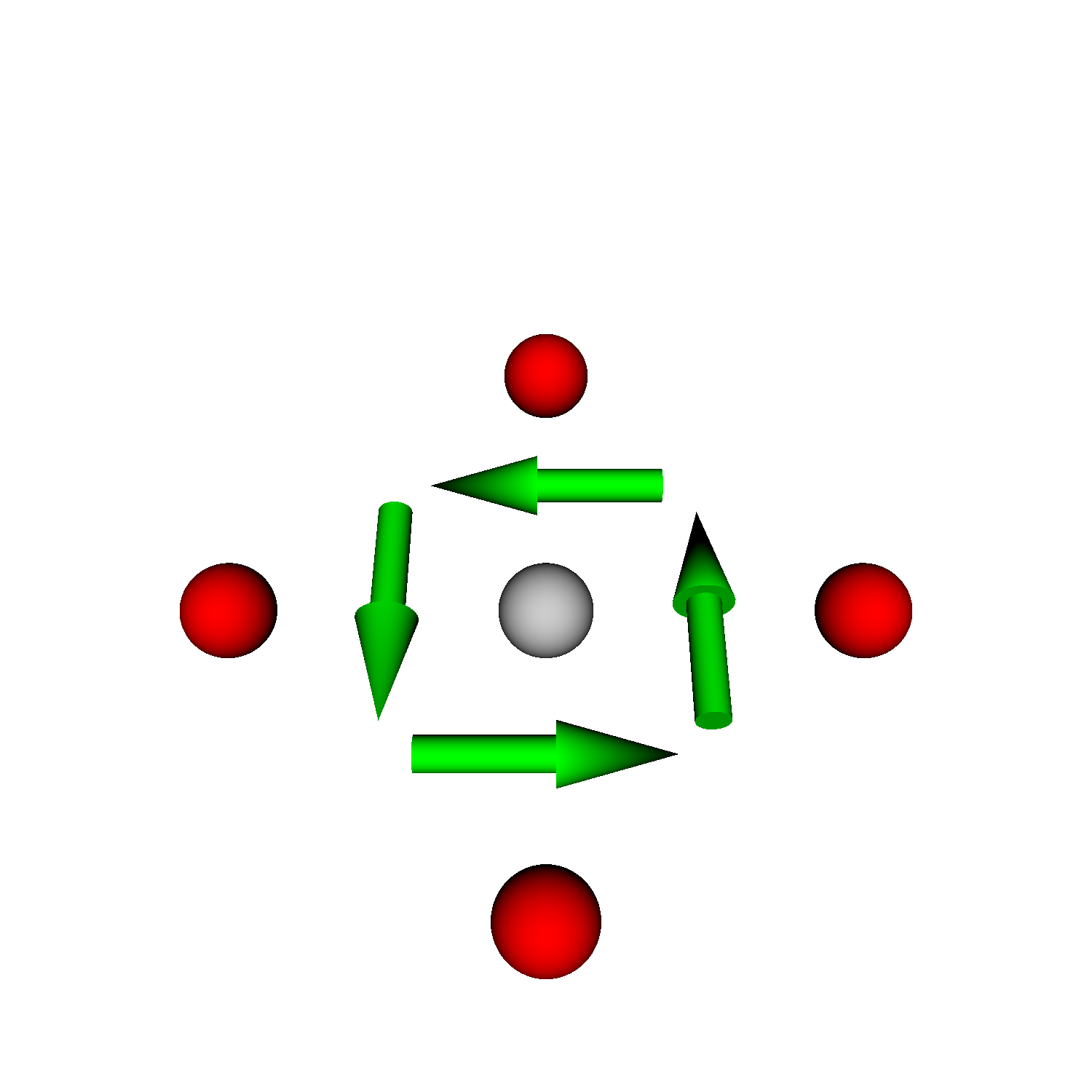} }
      \put(9.0, 0 ){ \includegraphics[height=4 cm, trim={0cm 0cm 0cm 0cm},clip=true]{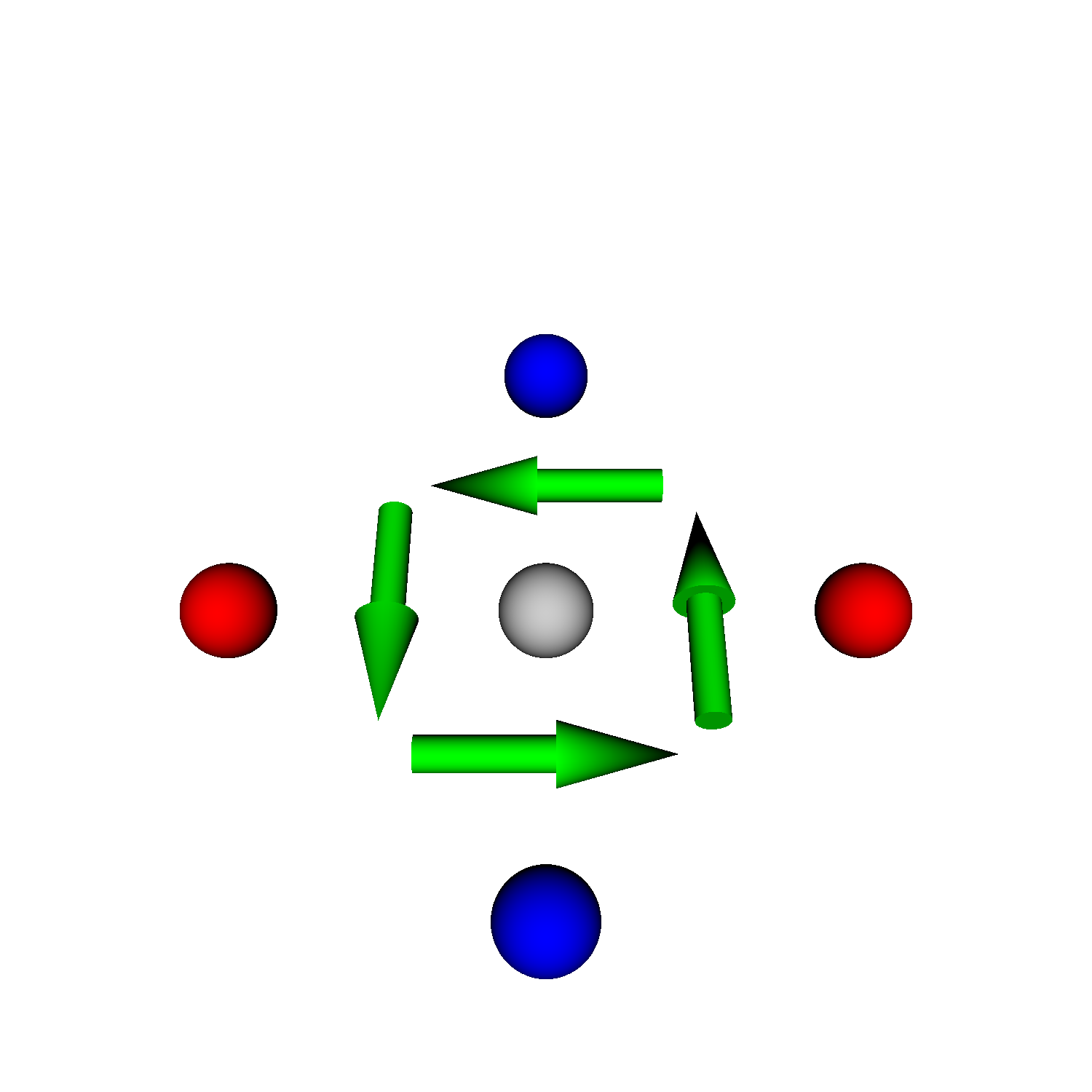} }
    }
    
    \put(-0.2,6.3){ \makebox(1,1){ (c) } } 
    \put( 7.3,6.3){ \makebox(1,1){ (d) } } 
    \put(-0.2,2.7){ \makebox(1,1){ (e) } } 
    \put( 7.3,2.7){ \makebox(1,1){ (f) } } 

    \put(0.2,3.75){ \framebox(0.75,0.75) } 
    \put(7.6,3.75){ \framebox(0.75,0.375) } 

    \newcommand{\boxsize}{0.3}
    \newcommand{\axisposX}{-0.2}
    \newcommand{\axisposY}{0.2}
    \put(\axisposX,\axisposY){ \vector(1,0){1}}
    \put(\axisposX,\axisposY){ \vector(0,1){1}}
    \put(\axisposX,\axisposY){ \vector(1,3){0.2}}
    \put(0.8,0.05){ \makebox(\boxsize,\boxsize){x} } 
    \put(-0.1,0.9){ \makebox(\boxsize,\boxsize){y} } 
    \put(-0.35,1.25){ \makebox(\boxsize,\boxsize){z} } 

  \end{picture}

  \caption{\label{fig:Fig_AF_spinconfig}
  Two-dimensional antiferromagnetic structures on a square lattice.
  (a) and (b) represent Model I and II, respectively.
  The gray atoms interact antiferromagnetically with the nearest-neighbor atoms in red, and ferromagnetically with the ones in blue.
  Both models also include an easy-axis magnetic anisotropy along $z$ (normal to the lattice plane).
  (c) and (d) show the c($2\times2$) and p($2\times1$) phases, which are the ground states of Model I and II without DMI, respectively.
  (e) and (f) show spin spirals formed due to Dzyaloshinskii-Moriya interactions in Models I and II, respectively.
  The energy per atom of the configuration in (c) and (d) is  $E=-4.050 J$, and in (e) and (f) $E=-4.052 J$.
  Model parameters: $D=0.2J$, $K=0.05J$ and $B=0$.
  }
\end{figure}
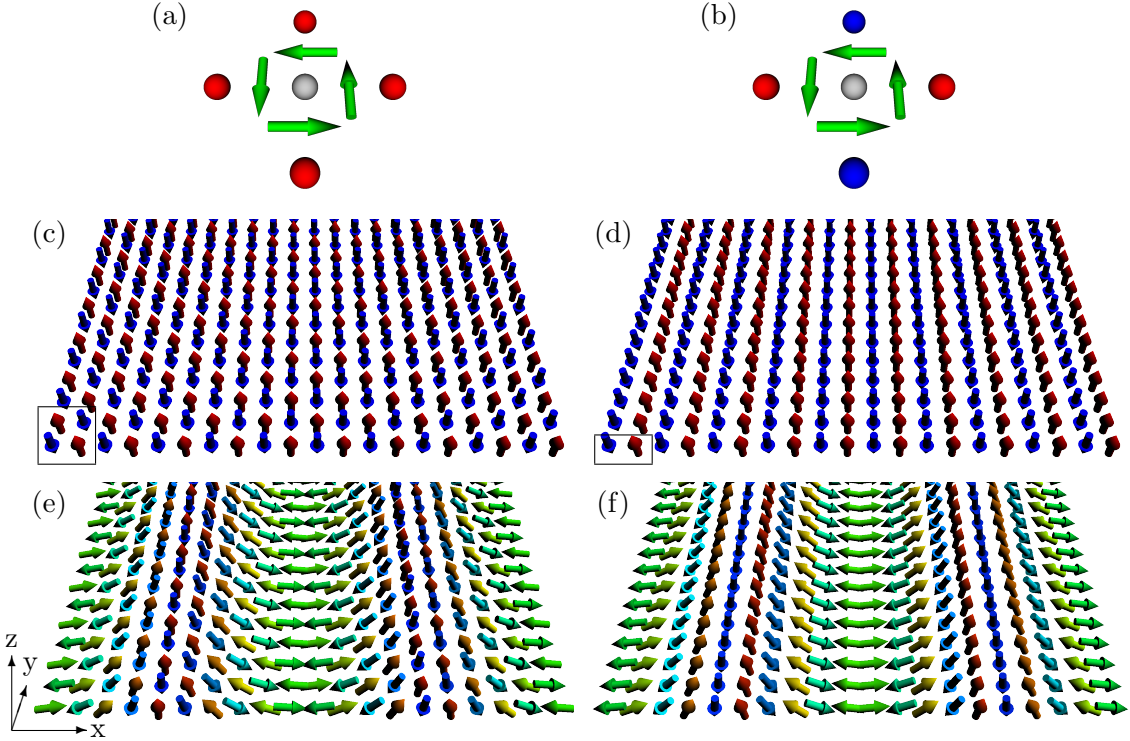

When the magnetic exchange interaction between neighboring atoms is dominant and antiferromagnetic, the formation of collinear structures of antiparallel spins takes place.
Considering a two-dimensional square lattice, the most common ones can be labelled according to the notation for surface reconstructions as the c($2\times2$) and p($2\times1$), shown in Figs.~\ref{fig:Fig_AF_spinconfig}(c) and (d).
In the c($2\times2$) structure, a spin moment is antialigned to all its nearest neighbors.
In the p($2\times1$), spins are antiparallel to their nearest neighbors along a given direction but align parallel along the perpendicular direction.

Here, we consider two related models with only nearest-neighbor interactions between spins on a square lattice.
In Model I, all nearest-neighbor exchange interactions are antiferromagnetic, which has the c($2\times2$) structure as the ground state, see Figs.~\ref{fig:Fig_AF_spinconfig}(a) and (c).
In an isotropic medium, beyond nearest-neighbor interactions are required to stabilize the p($2\times1$) structure~\cite{ferriani_magnetic_2007}.
To stick to only nearest-neighbor interactions, we circumvent this obstacle with a spatially anisotropic exchange interaction in Model II, such that $J_{ij} = -J$ along the $y$ and $J_{ij} = +J$ along the $x$--directions, as illustrated in Fig.~\ref{fig:Fig_AF_spinconfig}(b).
Its ground state, the p($2\times1$) structure, is shown in Fig.~\ref{fig:Fig_AF_spinconfig}(d).
The respective low-energy spin excitations have wavevectors around the $\mathrm{M}$--point for the c($2\times2$) structure and around the $\mathrm{X}$--point for the p($2\times1$) structure.
The $\mathrm{M}$--point $(\pi, \pi)$ corresponds to a precession where a given spin is dephased by $\pi$ with respect to all its nearest-neighbors.
For the c($2\times2$) structure, this means that nearest-neighbor spins are kept perfectly antiparallel throughout the whole precession revolution.
Such an excitation, therefore, costs no energy and corresponds to the Goldstone mode.
Similarly, the $\mathrm{X}$--point $(\pi, 0)$ guarantees a precession phase of $\pi$ between spins along horizontal lines, and no phase shift along the vertical ones, minimizing the excitation energy for the p($2\times1$) structure.

Since we want to study whether chiral skyrmions are supported by these antiferromagnetic systems, we need to consider the effects of the Dzyaloshinskii-Moriya interactions.
Thus, we add to both models in-plane isotropic nearest-neighbor Dzyaloshinskii-Moriya vectors, which circulate counterclockwise perpendicularly to the bonds.
Furthermore, both models include an easy-axis magnetic anisotropy favoring the spins to align along $z$.
The DMI favors a noncollinear alignment among the spins, while the anisotropy defines a preferred direction for the spin to point along.
The competition between the DMI, the magnetic anisotropy, and the exchange interaction determines the characteristics of the possible noncollinear states, such as spin spirals and skyrmions.

The ground states for Models I and II with DMI are shown in Fig.~\ref{fig:Fig_AF_spinconfig}(e) and (f), respectively, which we shall call c($2\times2$) and p($2\times1$)-spirals.
They were obtained by considering a $20\times2$ supercell for Model I, and a $20\times1$ supercell for Model II.
Notice that in the structures c($2\times2$), p($2\times1$), and p($2\times1$)-spiral there are always certain directions where the spins are aligned ferromagnetically.
Only the c($2\times2$)-spiral does not present such a feature.
Instead, we observe along its diagonals a smooth spin rotation forming a helical spin spiral.
The wavevector of both spin spirals is $\VEC q = q\,\hat{\VEC x}$ with magnitude $q = \pi/10$.

\begin{figure}[t]
\setlength{\unitlength}{1cm} 
\newcommand{\boxsize}{0.3}

  \begin{picture}(15.5,10)

    \put(0.2, 5){ \includegraphics[height=5. cm,trim={0cm 0 0 0},clip=true]{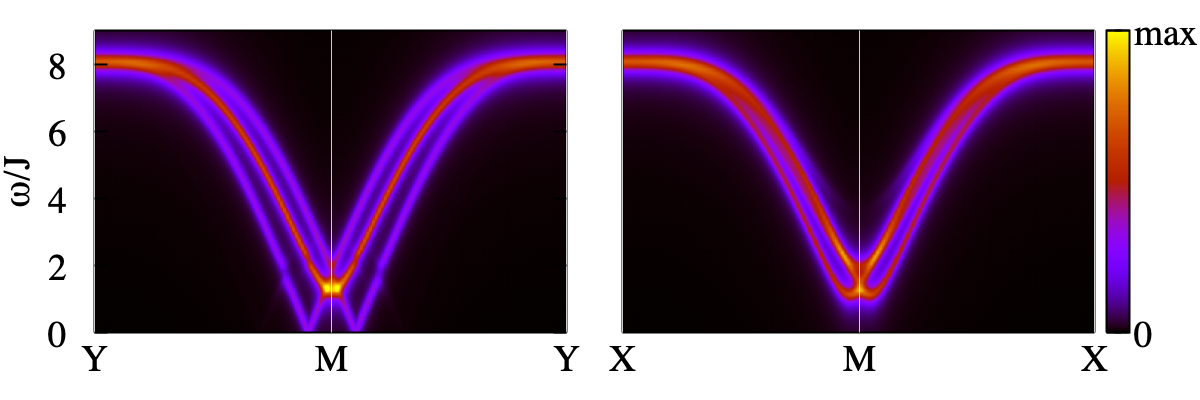} }
    \put(0.2, 0){ \includegraphics[height=5. cm,trim={0cm 0 0 0},clip=true]{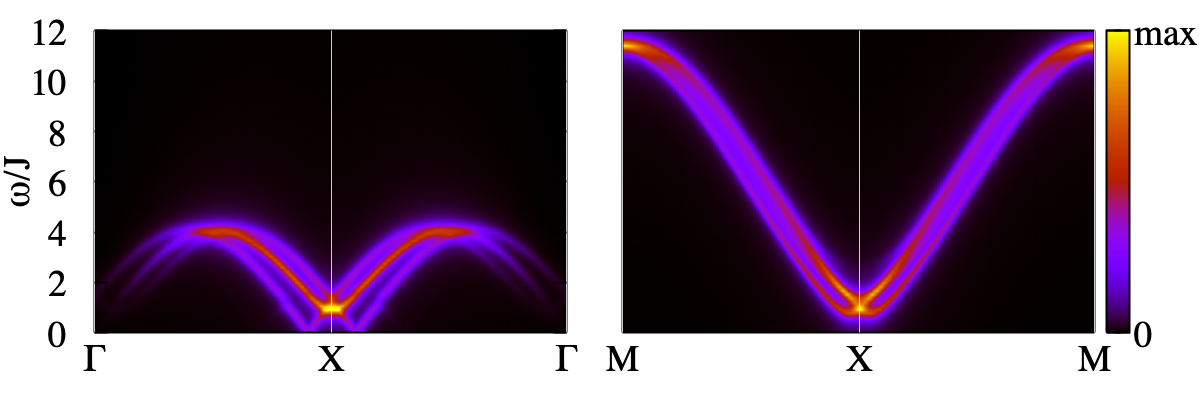} }

    \put(0.0, 9.4){ \makebox(\boxsize,\boxsize){(a)} } 
    \put(0.0, 4.6){ \makebox(\boxsize,\boxsize){(b)} }

    \put(2.3,2.8) {
      \color{white}
      \put(0.0,0.0){ \framebox(1.0,1.0) }  
      \multiput(0.58,0.5)(0.1,0.0){5}{\line(1,0){0.05}}
      \multiput(0.58,0.5)(0.0,0.1){5}{\line(0,1){0.05}}
      
      \put(0.45,0.5){ \circle*{0.08}}
      \put(0.45,1.0){ \circle*{0.08}}
      \put(0.95,0.5){ \circle*{0.08}}
      \put(0.95,1.0){ \circle*{0.08}}

      \put(1.0 ,0.32){ \makebox(0.4,0.4){X} }
      \put(1.0 ,1.1 ){ \makebox(0.4,0.4){M} }
      \put(0.15,1.1 ){ \makebox(0.4,0.4){Y} }
      \put(0.15,0.32){ \makebox(0.4,0.4){$\Gamma$} } 
    }
  \end{picture}

  \caption{\label{fig:Fig_AF_SP_disp}
  Inelastic scattering spectra for 2D antiferromagnetic spin spirals.
  (a) and (b) show the dispersion curves for the c($2\times2$) and p($2\times1$)-spirals, respectively.
  Paths along the spin-spiral wavevector $\VEC q$ are shown in the left-hand side and paths perpendicular to it on the right-hand side.
  The inset in (b) depicts the high symmetry points of the Brillouin zone for the underlying square lattice (the scattering process unfolds the spectra).
  Both the c($2\times2$) and p($2\times1$)-spirals feature the three universal helimagnon bands, seen on the left-hand side.
  Model parameters: $D=0.2J$, $K=0.05J$.
  }
\end{figure}
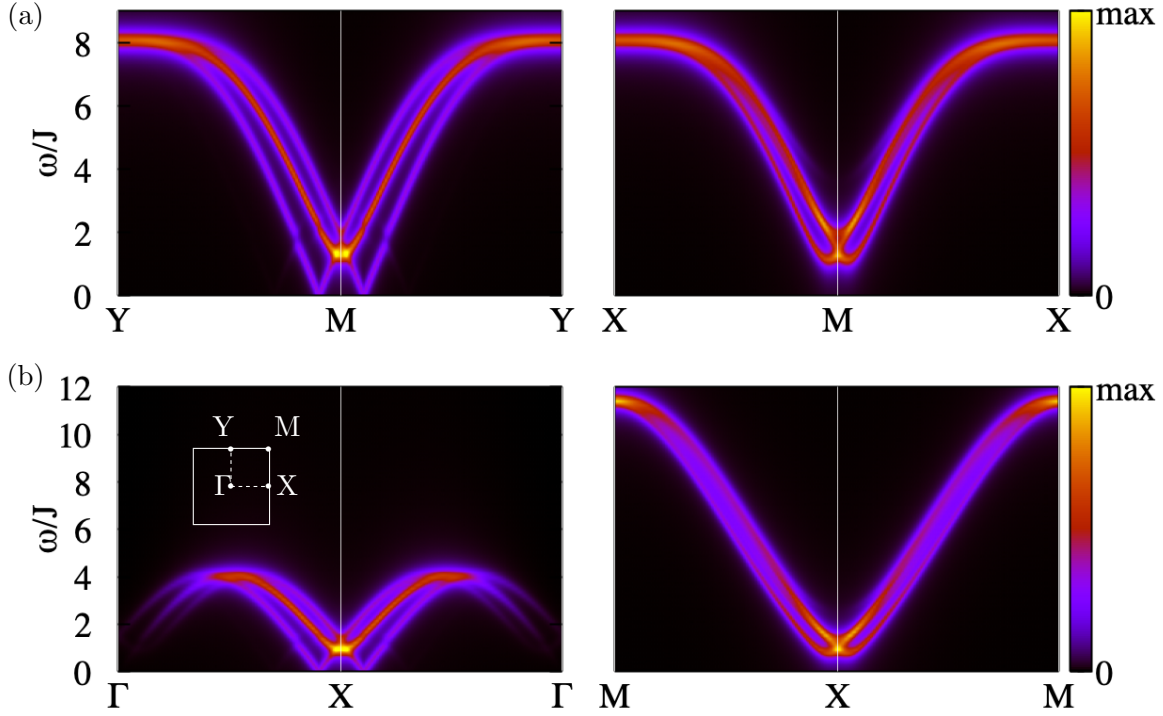

Next, we focus on the spin-wave spectra of these antiferromagnetic spin spirals.
In Fig.~\ref{fig:Fig_AF_SP_disp}(a), we show the total inelastic scattering spectrum for Model I in the c($2\times2$)-spiral state.
Two different reciprocal-space paths around the $\mathrm{M}$--point were considered, parallel and perpendicular to the spiral wavevector $\VEC q$.
Similarly, Fig.~\ref{fig:Fig_AF_SP_disp}(b) displays the total spectrum around the $\mathrm{X}$--point for Model II in the p($2\times1$)-spiral state.
Both spectra resemble each other, and display features that we already encountered for the one-dimensional antiferromagnetic spin spiral in Sec.~\ref{sec:ass}.
Namely, the three modes that are seen on the two left panels of Fig.~\ref{fig:Fig_AF_SP_disp} are precisely the two gapless rotational spin-wave modes dispersing away from the magnetic Bragg peaks of the spin spiral at $\mathrm{M} (\mathrm{X})\pm\VEC q$, while the central and more intense one is the longitudinal mode gapped by the magnetic anisotropy.
The energy scale of the spin waves of the p($2\times1$)-spiral along $\VEC q$ is roughly half of that for the c($2\times2$)-spiral.
The spin waves propagating along the $x$--axis have polarization along $y$, while the ones dispersing along $y$ are polarized along $x$.
These observations uncover a locking between the linear and angular momenta of the spin waves,
which is another characteristic of the Rashba effect discussed in Sec.~\ref{sec:effect_of_DMI_rashba}.

\begin{figure}[t]
\setlength{\unitlength}{1cm} 
\newcommand{\boxsize}{0.3}

  \begin{picture}(16,4.5)
    \put(0, 0){ \includegraphics[width=16 cm,trim={0.0cm 1cm 0 0},clip=true]{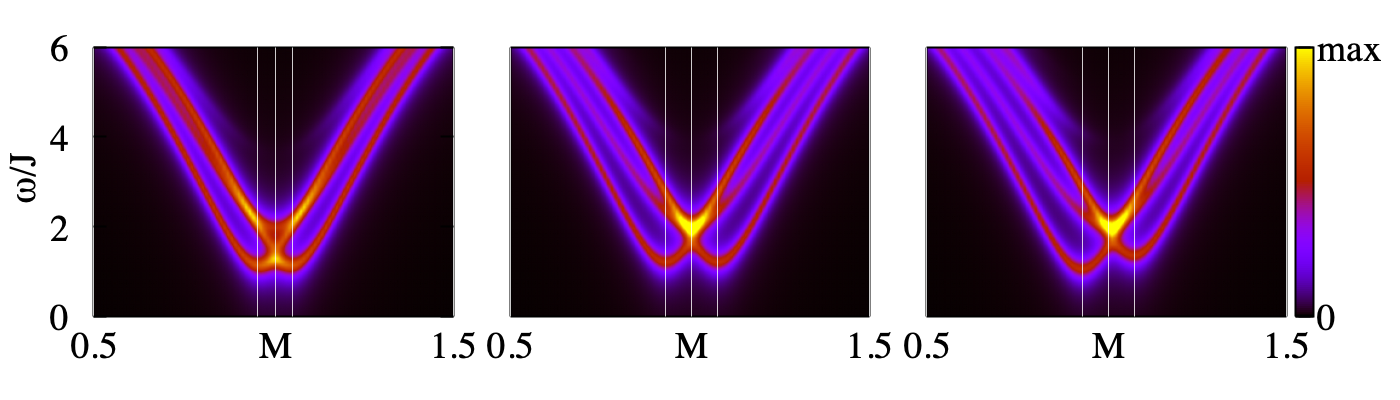} }

    \put( 1.1, 4.1){ \makebox(\boxsize,\boxsize){(a)} } 
    \put( 5.9, 4.1){ \makebox(\boxsize,\boxsize){(b)} }
    \put(11.1, 4.1){ \makebox(\boxsize,\boxsize){(c)} }
  \end{picture}

  \caption{\label{fig:Fig_AF_2d_tests}
  Inelastic scattering spectra along the X-M-X--path for the Model I on the c($2\times2$)-spiral, whose parameters were modified as follows.
  We increased the DMI from (a) $D=0.2J$ to (b) $D=0.3J$ without relaxing the spin structure ($B=0$).
  In (a) the energy minima are located at 
  $\mathrm{M} \pm 0.048$, while for (b) they are at $\mathrm{M} \pm 0.072$. Wavectors are given in unit of $\pi$.
  Thus the scaling on $D$ is linear.
  Due to the further splitting, a third mode can be distinguished, which is centered at $\mathrm{M}$.
  (c) Next, we apply an external magnetic field $B=0.2J$ along $\VEC q$ to the case in (b).
  The spectrum becomes nonreciprocal, with the magnetic field raising the excitation energies of one mode and lowering those of the other.
  Parameter: $K = 0.05$.
  }
\end{figure}
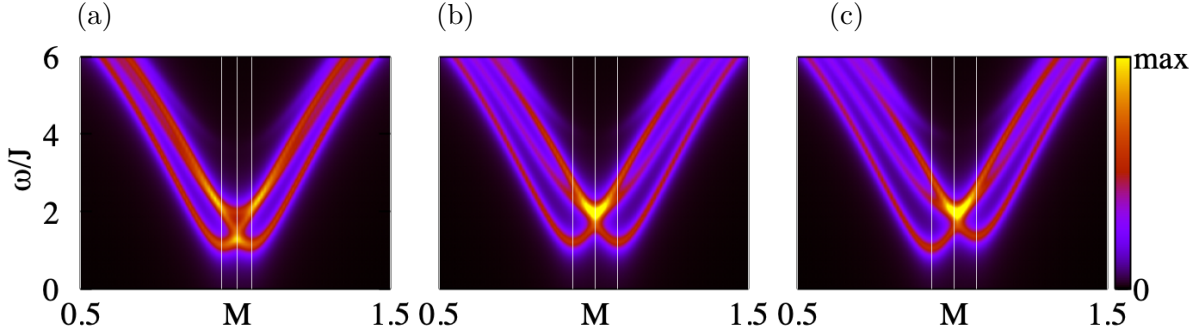

On the paths perpendicular to the spin-spiral wavevector (panels on the right-hand side in Fig.~\ref{fig:Fig_AF_SP_disp}), we initially observe two modes whose energy minima are shifted from the high-symmetry point.
Interestingly, the location in the reciprocal space of these minima is not related to the spin-spiral wavevector but directly to the strength of the Dzyaloshinskii-Moriya interaction in a linear manner.
We demonstrate this in Fig.~\ref{fig:Fig_AF_2d_tests}(a) and (b), where we increased the DMI strength from $D = 0.2J$ to $D = 0.3J$ without relaxing the spin structure, which resulted in a change of the minima from $\mathrm{M}\pm 0.048$ to $\mathrm{M} \pm 0.072$ ($\pi$).
Furthermore, the larger splitting induced by the enhanced DMI reveals a third mode that was indistinguishable before, see also Fig.~\ref{fig:Fig_AF_SP_disp}(a) and (b) (right-hand-side panels).
Finally, Fig.~\ref{fig:Fig_AF_2d_tests}(c) demonstrates that the two DMI-shifted modes are susceptible to external magnetic fields.
An applied field along the $x$--axis, therefore parallel to the polarization of these modes,
breaks the reciprocal symmetry around the band center, increasing the energy of one mode while decreasing the energy of the other one.
The inelastic scattering spectra of ferromagnetic and antiferromagnetic spin spirals have some similarities, such as the characteristic three helimagnon branches.
However, only the antiferromagnetic case displays DMI-shifted branches in all reciprocal space directions.

\subsection{Antiferromagnetic skyrmions}

\begin{figure}[t]
\setlength{\unitlength}{1cm} 

\newcommand{\figheight}{7}
\newcommand{\figlocy}{6.4}

\newcommand{\boxsize}{0.3}
  \begin{picture}(14,10.4)
    \put(-0.3,3.4){
        \put(0.3, 0){ \includegraphics[height=\figheight cm,trim={0 0 0 0},clip=true]{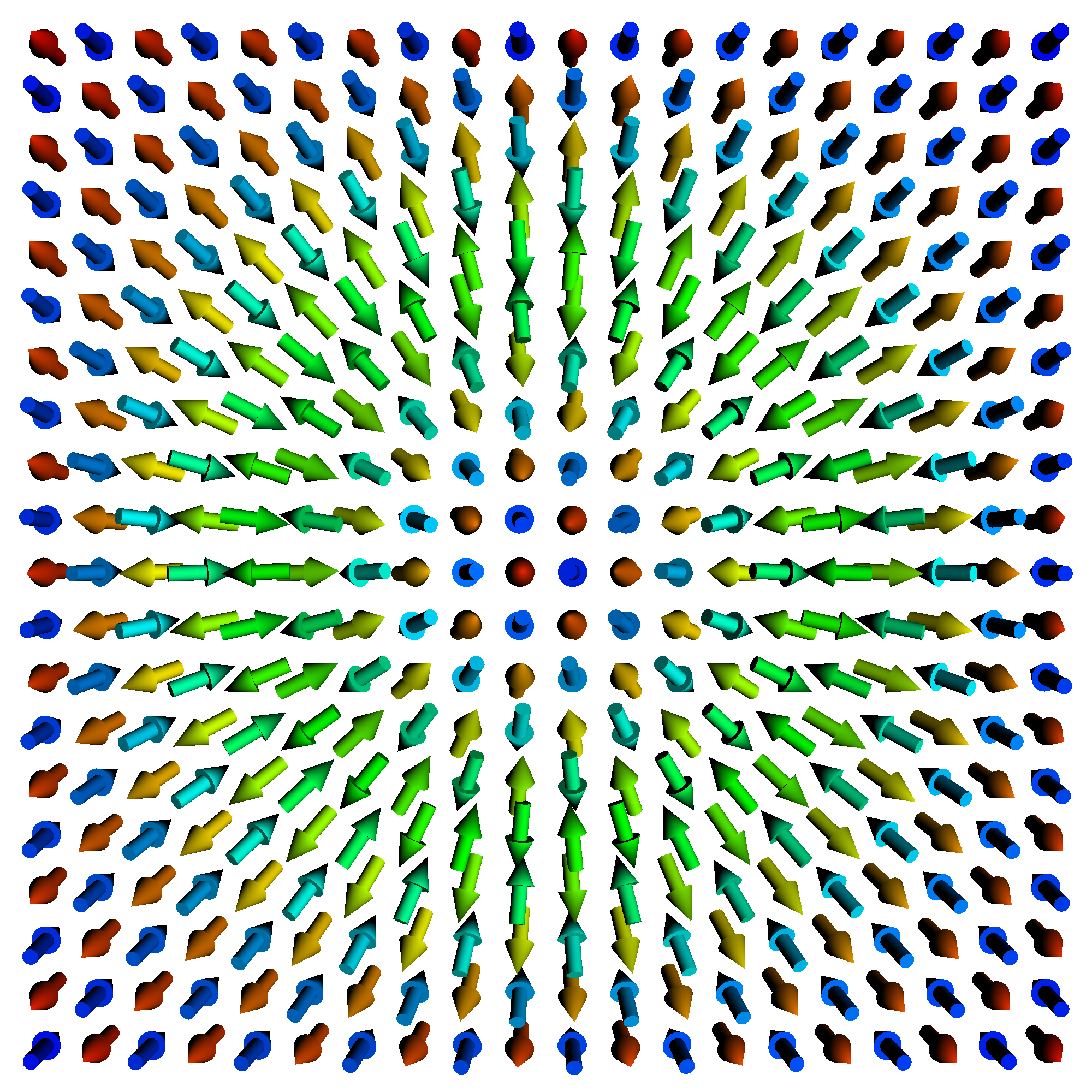} }
        \put(7.8, 0){ \includegraphics[height=\figheight cm,trim={0 0 0 0},clip=true]{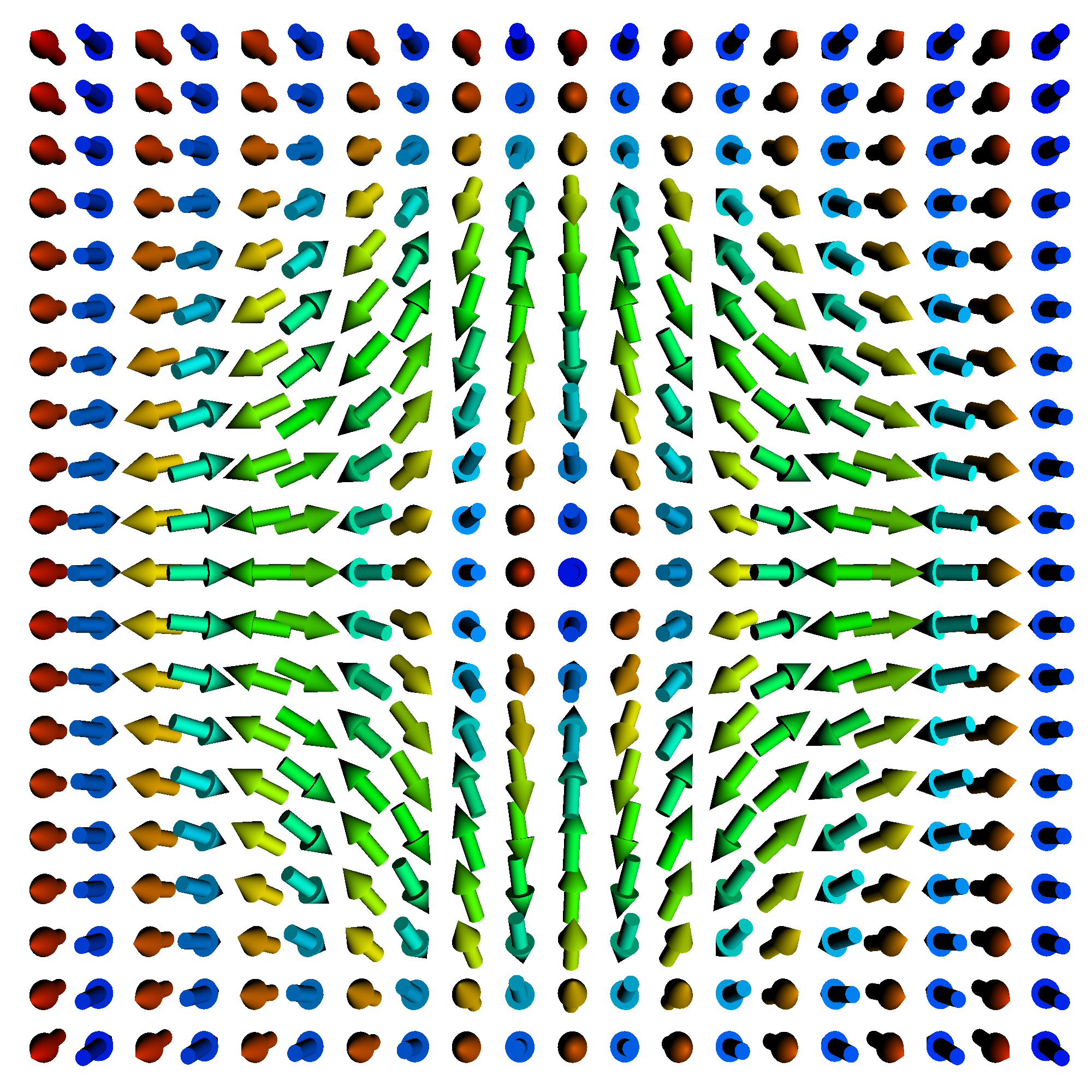} }
    
        \put(0,  \figlocy){ \makebox(\boxsize,\boxsize){(a)} } 
        \put(7.5,\figlocy){ \makebox(\boxsize,\boxsize){(b)} } 

        \put(0.2,-0.1){
          \put(0.0,0.0){ \vector(1,0){1} }
          \put(0.0,0.0){ \vector(0,1){1} }
          \put( 1.05,-0.15){ \makebox(0.3,0.3){x} }
          \put(-0.3, 0.9){ \makebox(0.3,0.3){y} }
        }
    }
    
    \put(1.0,0){
        \put(0.3, 0){ \includegraphics[width=12 cm,trim={0 0 0 0},clip=true]{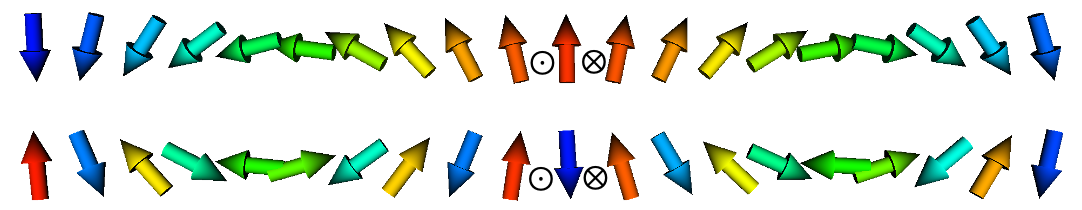} }
        
        \put(-0.5,  2.){ \makebox(\boxsize,\boxsize){(c)} } 
        
        \put(0.3, 1.2){
          \put(0.0,0.0){ \vector(1,0){0.8} }
          \put(0.0,0.0){ \vector(0,1){0.8} }
          \put( 0.85,-0.15){ \makebox(0.3,0.3){y} }
          \put(-0.3, 0.7){ \makebox(0.3,0.3){z} }
        }
        \put(0.3,-0.1){
          \put(0.0,0.0){ \vector(1,0){0.8} }
          \put(0.0,0.0){ \vector(0,1){0.8} }
          \put( 0.85,-0.15){ \makebox(0.3,0.3){x} }
          \put(-0.3, 0.7){ \makebox(0.3,0.3){z} }
        }
    }
  \end{picture}

  \caption{\label{fig:Fig_AF_SK}
  Skyrmionic structures in antiferromagnetic backgrounds.
  (a) Antiferromagnetic skyrmion that is formed when the exchange interaction with all nearest neighbors is negative, Model I.
  The skyrmion lives in a c($2\times2$) antiferromagnetic background.
  (b) Antiferromagnetic antiskyrmion, which results from spatially anisotropic exchange interactions.
  $J$ is negative along-$x$ and positive along-$y$, Model II.
  The antiskyrmion lies in a p($2\times1$) antiferromagnetic background.
  (c) Cross-sections along the $x$ and $y$ directions of the antiferromagnetic antiskyrmion in (b).
  They correspond to spin spirals with different chirality despite the same DMI, whose orientation is represented by the black circles.
  Model parameters: $D=0.2J$, $K=0.05J$ and $B=0$. 
  The total energy of both spin configuration is $E=-4.041 J$.
  }
\end{figure}
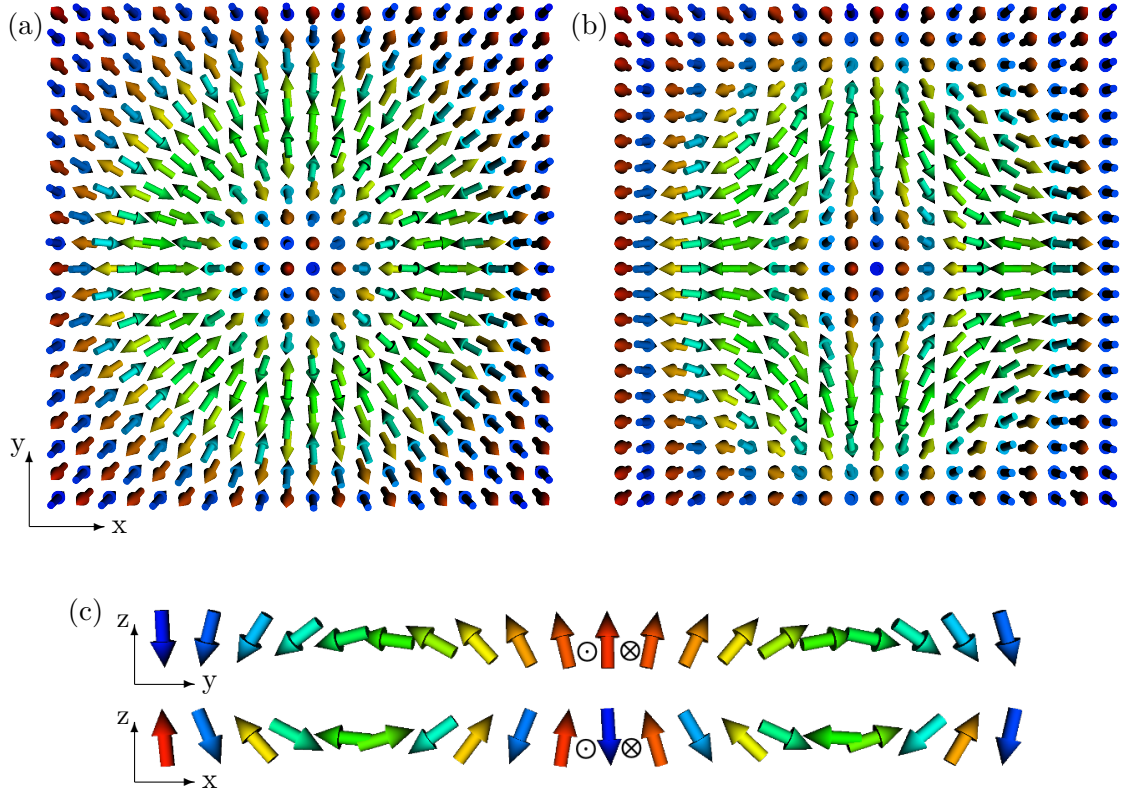

In the previous section, we considered two model Hamiltonians with c($2\times2$) and p($2\times1$) antiferromagnetic spin spirals as ground states.
Next, we performed atomistic-spin-dynamics calculation using a square simulation box matching the wavelength of those spin spirals with periodic boundary conditions.
Thus, we obtain antiferromagnetic skyrmionic lattices as metastable configurations of Models I and II.
On the one hand, Model I, whose atoms have an antiferromagnetic exchange interaction with all their nearest neighbors, gives rise to an antiferromagnetic skyrmion, as can be seen in Fig.~\ref{fig:Fig_AF_SK}(a).
On the other hand, Model II stabilizes an antiferromagnetic antiskyrmion, see Fig.~\ref{fig:Fig_AF_SK}(b), which has an anisotropy profile around the core in contrast with the isotropic profile of the antiferromagnetic skyrmion.
Here, it is important to notice that Model I and II share the same set of isotropic DMI.
Their only difference lies in the set of exchange interactions, where for Model II the exchange parameter $J$ changes sign for different directions.
For a given set of DMI, the antiferromagnetic alignment reverses the chirality of the spin spirals in comparison to the chirality of a ferromagnetic arrangement, see Fig.~\ref{fig:Fig_AF_SK}(c).
As the p($2\times1$) state has a ferromagnetic cross-section along $y$ and an antiferromagnetic cross-section along $x$, the antiskyrmion is the natural occurrence for this type of antiferromagnets.
We also confirmed this result by employing a next-nearest-neighbor isotropic Hamiltonian with $C_{4v}$ symmetries that favors the p($2\times1$) state through exchange frustration, therefore, without invoking a spatially anisotropic magnetic exchange interaction.

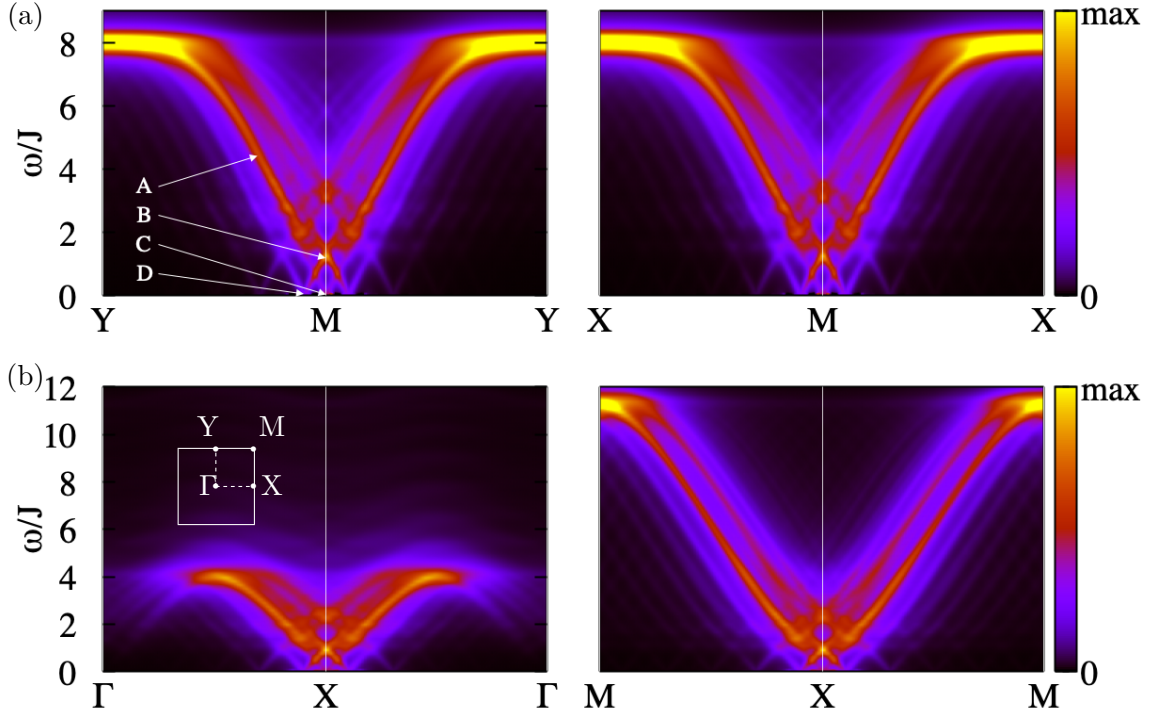
\begin{figure}[t]
\setlength{\unitlength}{1cm} 
\newcommand{\boxsize}{0.3}
  \begin{picture}(14,10)
    \put(-0.1, 5){ \includegraphics[height=5. cm,trim={0cm 0 0 0},clip=true]{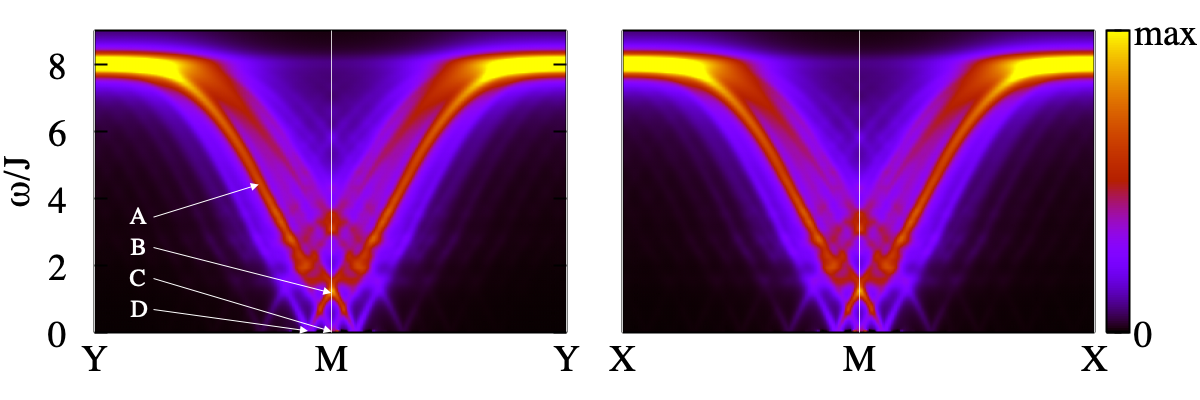} }
    \put(-0.1, 0){ \includegraphics[height=5. cm,trim={0cm 0 0 0},clip=true]{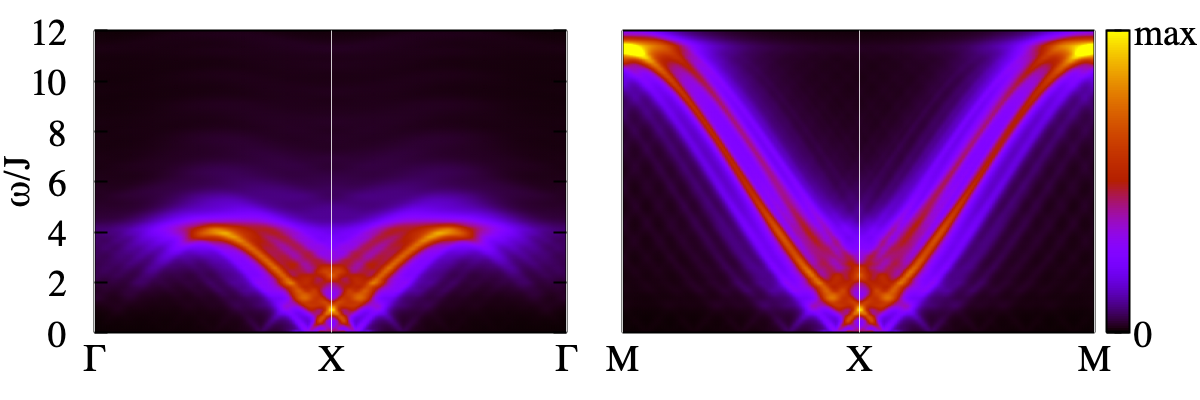} }

    \put(-0.1, 9.4){ \makebox(\boxsize,\boxsize){(a)} } 
    \put(-0.1, 4.6){ \makebox(\boxsize,\boxsize){(b)} }

    \put(2.0,2.8) {
      \color{white}
      \put(0.0,0.0){ \framebox(1.0,1.0) }  
      \multiput(0.58,0.5)(0.1,0.0){5}{\line(1,0){0.05}}
      \multiput(0.58,0.5)(0.0,0.1){5}{\line(0,1){0.05}}
      
      \put(0.45,0.5){ \circle*{0.08}}
      \put(0.45,1.0){ \circle*{0.08}}
      \put(0.95,0.5){ \circle*{0.08}}
      \put(0.95,1.0){ \circle*{0.08}}

      \put(1.0 ,0.32){ \makebox(0.4,0.4){X} }
      \put(1.0 ,1.1 ){ \makebox(0.4,0.4){M} }
      \put(0.15,1.1 ){ \makebox(0.4,0.4){Y} }
      \put(0.15,0.32){ \makebox(0.4,0.4){$\Gamma$} } 
    }

  \end{picture}

  \caption{\label{fig:Fig_AF_SK_disp}
  Inelastic scattering spectra for antiferromagnetic skyrmionic lattices.
  (a) Spectrum for the antiferromagnetic skyrmion in the c($2\times2$) background.
  (b) Spectrum for the antiferromagnetic antiskyrmion in the p($2\times1$) background.
  In contrast to the spin spirals, the intense features for the skyrmion lattices are much more broadened.
  The inset in (b) depicts the high symmetry points of the crystal Brillouin zone.
  Model parameters: $D=0.2J$, $K=0.05J$ and $B=0$.
  }
\end{figure}

Next, we investigate the inelastic scattering spectra related to these skyrmionic structures.
Fig.~\ref{fig:Fig_AF_SK_disp}(a) shows the spectra for the antiferromagnetic skyrmion, which is related to the c($2\times2$) antiferromagnetic structure, while 
Fig.~\ref{fig:Fig_AF_SK_disp}(b) corresponds to the antiferromagnetic antiskyrmions, whose background is a p($2\times1$) antiferromagnet.
Overall, both spectra have a lot more structure in comparison to those of the parent antiferromagnetic spin spirals, with many faint modes almost forming a continuum of spin-wave excitations.
Yet, it is possible to clearly resolve distinct intense modes throughout most of the reciprocal path.

For small excitation energies around $\mathrm{M}$ or $\mathrm{X}$, we can observe that the modes disperse mostly linearly with the changing wavevector.
This is in contrast to the quadratic behavior of the low-energy spin-wave modes for ferromagnetic skyrmion lattices~\cite{dos_santos_spin-resolved_2018}.
In Fig.~\ref{fig:Fig_AF_SK}(a), we singled out a few points in the spectrum with features common to all four panels.
The arrow labeled A indicates a continuous dispersing mode which is circularly polarized with net angular momentum along $+y$ on the left half of the $\mathrm{Y}-\mathrm{M}-\mathrm{Y}$ path.
The same feature on the right-hand side has angular momentum along $-y$.
The polarization of this mode rotates to the $x$ direction in the $\mathrm{X}-\mathrm{M}-\mathrm{X}$ path, for example.
The B arrow indicates a hotspot seen in all four panels with nonvanishing energy and localized at the high symmetry points.
It corresponds to a mode with linear polarization along $z$.
The intensity marked by the C arrow is due to two degenerate rotational modes with opposite angular moment along $z$, therefore, reproducing the polarization of the native modes of the collinear antiferromagnetic background.
The modes dispersing out of the Bragg peaks, such as the one indicated by D, are linearly polarized along $z$.
The precession nature and relative energy alignment of the excitations due to points B and C are in accordance with the ones reported for antiferromagnetic skyrmions confined in nanodiscs~\cite{Kravchuk2019}.

Notice that all the spectra are symmetric with respect to the high-symmetry points.
As the net magnetization is zero and no magnetic field has been applied, the spin-wave energies must be reciprocal.
Nevertheless, hidden nonreciprocity of individual spin-wave modes induced by the DMI could be observed as an asymmetry in the inelastic scattering spectra if spin-polarized spectroscopies are to be used~\cite{dos_santos_nonreciprocity_2020}.

\section{Conclusions}

We studied simple models of magnetic materials whose magnetic exchange interaction is predominately antiferromagnetic, intending to contribute to the development of antiferromagnetic spintronics and magnonics.
We first considered one-dimensional antiferromagnetic spin chains.
We observed that the DMI can lift the degeneracy of the two spin-wave modes in the collinear antiferromagnetic structure, resulting in a mode splitting similar to the Rashba effect for electronic bands.
Because these two spin-wave modes have opposite angular momenta, a magnetic field can induce a nonreciprocity of the spectrum.
For even higher fields, the collinear state gives way to a spin-flop state where a spin spiral is formed.
The new state has a reciprocal spin-wave spectrum formed of the three universal helimagnon modes~\cite{garst_collective_2017}.
Our calculations compare well with the recent experimental results obtain with inelastic neutron scattering for the bulk material $\alpha$--$\textnormal{Cu}_2\textnormal{V}_2\textnormal{O}_7$~\cite{gitgeatpong_nonreciprocal_2017}, and explain some unremarked features in the measurements.

We also investigated noncollinear spin textures in two dimensions.
In particular, we computed the inelastic scattering spectra for two model systems with spin-spiral ground states, one with spatially-isotropic and another with spatially-anisotropic magnetic exchange interactions sharing the same set of isotropic DMI.
The isotropic interactions favor the c($2\times2$) checkerboard antiferromagnetic structure in the isotropic case and the p($2\times1$) row-wise antiferromagnetic structure in the anisotropic case, and the DMI leads to spiral structures based on those reference collinear states.
The spin-wave spectra have some similarities with those for ferromagnetic spin spirals studied in Ref.~\cite{dos_santos_spin-resolved_2018}, but are centered at high-symmetry points at the edges of the Brillouin zone instead of at the zone center.
The different modes were characterized in terms of their precessional character, which is tied to the chosen path in reciprocal space in a way that is once again reminiscent of the Rashba spin-momentum locking.
With the same models, we could also stabilize antiferromagnetic skyrmion lattices.
We demonstrated that antiskyrmions are the natural skyrmionic occurrence in p($2\times1$) antiferromagnets because the antiferromagnetic alignment imposes a chirality switch.
Remarkably, these antiskyrmions can be obtained with isotropic DMI, which is in contrast to ferromagnetic materials.
Lastly, we calculated the inelastic scattering spectra for the antiferromagnetic skyrmion and antiskyrmion lattices and identified breathing and gyroscopic modes in the resulting spectra.

\begin{acknowledgments}

This work is supported by the Brazilian funding agency CAPES under Project No. 13703/13-7 and the European Research Council (ERC) under the European Union's Horizon 2020 research and innovation programme (ERC-consolidator Grant No. 681405-DYNASORE). We gratefully acknowledge the computing time granted by JARA-HPC on the supercomputer JURECA at Forschungszentrum Jülich and by RWTH Aachen University.
\end{acknowledgments}

\begin{appendix}

\section{Antiferromagnetic spin spirals}
\label{Apx:Stiff_spiral_tensors}

\subsection{Spin-spiral model}
\label{sub:spin_spiral}

Let us suppose that we have the magnetic properties of a given system.
That is, we have the spin moment of each site and the magnetic interactions, both the exchange and the Dzyaloshinskii-Moriya, from each pair of atoms in a Bravais lattice.
We want to determine whether a spin spiral can be an energetically more favorable state than a collinear antiferromagnetic phase.
However, there can be many types of spin spirals, with different orientations, wavevectors, etc.
Thus, we restrict our search among spirals given by the following equation:
\begin{equation}\label{eq:spin_spiral_DMI}
  \VEC S_i = \cos\phi_i \sin\theta\,\VEC n^1 + \sin\phi_i \sin\theta\,\VEC n^2 + \cos\theta\,\VEC n^3 \quad ,
\end{equation}
where $\VEC n^3$ is a unity vector defining the axis around which the spins rotate, and that forms an orthonormal basis set for the three dimension space together with $\VEC n^1$ and $\VEC n^2$.
$\theta$ is the conical angle between the spins and $\VEC n^3$.
$\phi_i = (\VEC q_\textnormal{AF} + \VEC q ) \cdot \VEC R_i $, where $\VEC q_\textnormal{AF}$ accounts for a collinear antiferromagnetic phase,  $\VEC q$ is the spiral wavevector, and $\VEC R_i$ the position vector of the $i$-th spin. 

\subsection{Spin-spiral energy} 
\label{sub:spin_spiral_energy}

Regarding the hamiltonian of Eq.~\eqref{eq:Hamiltonian}, the classical energy (per atom) of such a spin spiral can be decomposed in four terms.

The exchange one:
\begin{equation}
\begin{split}
  \varepsilon_J(\VEC q)
   =& \frac{1}{N}\sum_{ij} J_{ij} \VEC S_i \cdot \VEC S_j \\
   =& \frac{1}{N}\sum_{ij} J_{ij} \left[ \cos\big((\VEC q_\textnormal{AF} + \VEC q )\cdot \VEC R_{ij}\big) \sin^2\theta + \cos^2\theta \right] \quad .
\end{split} 
\end{equation}

The Dzyaloshinskii-Moriya term:
\begin{equation}
\begin{split}
 \varepsilon_D(\VEC q) = & -\frac{1}{N}\sum_{ij} \VEC D_{ij} \cdot \VEC{S}_i \times \VEC{S}_j \\
= &  -\frac{1}{N} \sum_{ij}  
      \sin^2\theta  D_{ij}^3 \sin\big( (\VEC q_\textnormal{AF} + \VEC q) \cdot \VEC R_{ij}\big) \quad ,
\end{split}
\end{equation}
where we used the symmetry property of the Dzyaloshinskii-Moriya interaction that imposes $\VEC D_{ij} = - \VEC D_{ji}$.

And the uniaxial magneto-crystalline anisotropy contribution:
\begin{equation}
\begin{split}
  \varepsilon_K(\VEC q)
  =& - \frac{K}{N} \sum_{i} ( \hat{\VEC K} \cdot \VEC S_i )^2 \\
  =& - K \left[ \frac{1}{2}\sin^2\theta \Big( (K^1)^2 (\delta_{\VEC q, \{0, \VEC q_\textnormal{AF}\}} + 1) + (K^2)^2 (1 - \delta_{\VEC q, \{0, \VEC q_\textnormal{AF}\}} ) \Big) \right.\\
  & \quad \qquad \left. + \sin 2\theta \big( K^1 K^3 \delta_{\VEC q, \VEC q_\textnormal{AF}} \big) + \cos^2\theta (K^3)^2  \right] \quad , \\
\end{split} 
\end{equation}
where we considered that $\VEC q$ is restricted to the first Brillouin zone.
The unit vector that represent the magnetic anisotropy axis decomposes as $\hat{\VEC K} = K^1 \VEC n^1 + K^2 \VEC n^2 + K^3 \VEC n^3$ with $\sum^\alpha ( K^\alpha )^2 = 1$.
Thus, we can see that the anisotropy energy is a constant for every wavevector different of zero:
\begin{equation}
\begin{split}
  \varepsilon_K(\VEC q \neq 0)
  =& - K \left\{ \frac{1}{2} \sin^2\theta \left[ (K^1)^2  + (K^2)^2 \right] + \cos^2\theta (K^3)^2  \right\} \quad, \\
\end{split} 
\end{equation}
and it contains a singularity for the antiferromagnetic state:
\begin{equation}
\begin{split}
  \varepsilon_K(\VEC q = 0)
  =& - K \left [ \sin^2\theta (K^1)^2  + \cos^2\theta (K^3)^2  \right] \quad. \\
\end{split} 
\end{equation}
The component $K^2$ does not appear in this last result because for $\VEC q = 0$ in the definition of the spin spiral, Eq.~\eqref{eq:spin_spiral_DMI}, the circular components of the spins point along $\VEC n^1$ only. 

The magnetic field term is given by:
\begin{equation}
\begin{split}
\varepsilon_B(\VEC q) = & - \sum_i \VEC B\cdot \VEC S_i \\
  =& - \sum_i  B^3 \cos\theta \VEC n^3  \quad ,
\end{split}
\end{equation}
where $B^\alpha = \VEC B \cdot \VEC n^\alpha$.

\subsection{One-dimensional model}

We now consider the one-dimensional model introduced in Sec.~\ref{sec:1d_chain}, with $\VEC D = D \hat{\VEC z}$, $\VEC K = \hat{\VEC z}$, and $\VEC B = B \hat{\VEC z}$.
Energy of the antiferromagnetic state with spin parallel to the anisotropy easy-axis reads
\begin{equation}
    \begin{split}
        \varepsilon_\textnormal{AF}
        = & -2J - K \quad .
    \end{split}
\end{equation}
Meanwhile, the energy  of the antiferromagnetic spin-spiral state with rotational axis along $z$ is
\begin{equation}
    \begin{split}
        \varepsilon_\textnormal{spiral}( q)
        = & -2J \left[\cos\big(a q\big) \sin^2\theta  - \cos^2\theta \right] 
        -2D \sin\big( a q\big)\sin^2\theta 
        - K \cos^2\theta
        - B \cos \theta \quad .
    \end{split}
\end{equation}
The spiral wavevector that minimizes the total energy is given by
\begin{equation}
    \begin{split}
        \frac{\partial\varepsilon_\textnormal{spiral}( q_\textnormal{min})}{\partial q}
        = & 2a\Big( J \sin\big(a q_\textnormal{min}\big)
        -D \cos\big( a q_\textnormal{min}\big)\Big)\sin^2\theta =  0 \\
     &\implies \tan\big(a q_\textnormal{min}\big)  =  \frac{D}{J}  \quad . \\
    \end{split}
\end{equation}
Thus, the mininum energy is achieved when the spin-spiral wavevector satisfy $\tan(a q_\textnormal{min}) = D/J$.
Replacing this result in the total energy, we have that the minimum energy is:
\begin{equation}
\begin{split}
  \varepsilon_\textnormal{spiral}(q_\textnormal{min})
  =& - 2 \sqrt{J^2 + D^2} \sin^2\theta - (-2 J + K ) \cos^2\theta - B \cos\theta \quad ,\\
\end{split} 
\end{equation}
where we used the transformation $\sin(a q) = D / \sqrt{J^2+D^2}$ and $\cos(a q) = J / \sqrt{J^2+D^2}$.

The spin spiral becomes more favorable when its energy is lower than the antiferromagnetic state energy thus satisfying:
\begin{equation}
\begin{split}
  \varepsilon_\textnormal{AF} - \varepsilon_\textnormal{spiral}(q_\textnormal{min}) >& 0 \\
 \Delta\varepsilon = \big(2 \sqrt{J^2 + D^2} - K + 2J \big) \sin^2\theta -4 J  + B \cos\theta   >& 0  \quad .\\
\end{split}
\end{equation}
The first term is larger than zero when $K < 2 \sqrt{J^2 + D^2} + 2J$, where it is maximized for $\theta = \pi/2$.
The term that depends on the magnetic field is maximized for $\theta = 0$.
For $B = 0$ and $\theta = \pi/2 $, the condition is satisfied when
\begin{equation}
\begin{split}
  D   >&  \sqrt{\left(J + \frac{K}{4}\right) K} \quad .\\
\end{split}
\end{equation}

Let us find the global maximum of the energy difference as a function of $\theta$:
\begin{equation}
\begin{split}
 \Delta \varepsilon (\theta) = &   \big(2 \sqrt{J^2 + D^2} - K + 2J \big) \sin^2\theta -4 J  + B \cos\theta  \\
 \frac{\partial}{\partial \theta}\Delta \varepsilon (\theta) = &   \big(2 \sqrt{J^2 + D^2} - K + 2J \big) 2\sin\theta \cos\theta  - B \sin\theta  \quad , \\
 \frac{\partial^2}{\partial \theta^2}\Delta \varepsilon (\theta) = &   \big(2 \sqrt{J^2 + D^2} - K + 2J \big) 2\big(2\cos^2\theta-1\big)  - B \cos\theta \quad .  \\
\end{split}
\end{equation}
The critical points are given by (we only need to check for $0\le \theta \le \pi$):
\begin{equation}
\begin{split}
  \theta_1 = 0 \quad, \quad   \cos\theta_2  = \frac{B}{2\big(2 \sqrt{J^2 + D^2} - K + 2J \big) } \quad .
\end{split}
\end{equation}
Analyzing the concavity of these points, we have:
\begin{equation}
\begin{split}
 \frac{\partial^2}{\partial \theta^2}\Delta \varepsilon (\theta_1) = &   \big(1  - \cos\theta_2 \big)  \frac{B}{\cos\theta_2} > 0 \quad ,  \\
 \frac{\partial^2}{\partial \theta^2}\Delta \varepsilon (\theta_2) = &   \big( \cos^2\theta_2 - 1\big) \frac{B}{\cos\theta_2} < 0  \quad ,  \\
\end{split}
\end{equation}
because $ 0 \le \cos\theta_2\le 1$.
Theferore, $\theta_2$ should be the global maximum with energy difference given by
\begin{equation}
\begin{split}
 \Delta \varepsilon (\theta_2) = &   \frac{B}{2 } \left(\frac{1}{ \cos\theta_2} + \cos\theta_2 \right) -4 J   \quad .
\end{split}
\end{equation}

\end{appendix}

\bibliography{Antiferro}

\end{document}